# Polyethylene-based thermo-mechanically recyclable stretchable yarns for circular sustainable textiles

SeongHyeon Kim[1,†], Duo Xu[1,†], Volodymyr Korolovych[3], Domingo R. Flores-Hernandez[1,2], Kaniz Moriam[1], Daniel J. Braconnier[1], and Svetlana V. Boriskina[1*]

[1] Massachusetts Institute of Technology, Department of Mechanical Engineering, Cambridge, MA, 02139, USA

[2] Tecnologico de Monterrey, Escuela de Ingenieria y Ciencias, Ave. Eugenio Garza Sada 2501, Monterrey 64849, Mexico

[3] Massachusetts Institute of Technology, Institute for Soldier Nanotechnologies, Cambridge, MA, 02139, USA

[†] These authors contributed equally to this work.

*Corresponding author e-mail: sborisk@mit.edu

**ABSTRACT**

Most high-performance elastic textiles rely on yarns composed of chemically dissimilar polymers, rendering them difficult to recycle. Here, we demonstrate fully thermo-mechanically recyclable stretchable yarns composed of polyethylene (PE) family materials. Inspired by structure-property relationships in natural materials, we engineer a library of melt-spun PE fibers spanning mechanical properties from elastomeric to functional by tuning polymer crystallinity and chain orientation. These fibers are assembled into core-sheath yarns comprising an olefin block copolymer elastic core and a high-strength PE sheath, forming a helical architecture. The resulting yarns exceed mechanical performance of commercial PET-spandex yarns while maintaining full recyclability. We further show that PE homopolymers and copolymers can be jointly melt-processed and recycled without phase separation or loss of performance. This approach enables stretchable recyclable textiles from fibers with previously demonstrated cooling, moisture-wicking and stain-resisting performance and provides a scalable pathway toward circular garments compatible with existing polyethylene recycling streams.

The textile and apparel industries carry a massive environmental footprint across all production stages, from high energy and water consumption during raw material synthesis and extraction, yarn spinning, and coloring to microplastic pollution during use and disposal [1,2]. Fast fashion exacerbates this problem, as most garments are discarded after a few uses and landfilled at a rate equivalent to one garbage truck per second [3]. Breaking this cycle requires functional, durable, fully recyclable textiles, i.e., materials with (i) a lower production footprint, (ii) durability through resistance to staining and degradation, and (iii) recyclability within existing waste systems [2,4–9].

A key barrier is that most garments remain virtually impossible to recycle thermo-mechanically because they contain inseparable polymer mixtures. Conventional polyurethane-based elastic fibers, such as spandex (elastane/Lycra) [10,11], exhibit low tensile strength and must be plied or core-spun with higher-tenacity yarns, such as polyester (PET) [12,13], cotton, wool, or nylon, to form functional elastic yarns. Although PET and nylon are recyclable, and cotton and wool are biodegradable, combining any with spandex requires material separation at the end of the lifecycle [2,10,14–16]. Existing recycling methods, such as solvent extraction or thermal degradation of spandex, rely on harsh chemicals and high energy inputs [17,18], making them unsustainable at scale.

Polymer fibers can also be incinerated or thermo-chemically recycled via pyrolysis. However, mixtures remain challenging due to dissimilar chemistries, differing process temperatures, and varied solid, liquid, and gaseous compounds [5,9,19]. Nitrogen in nylon, wool, and spandex generates harmful emissions requiring specialized treatment, while polyester pyrolysis requires specific catalysts to raise useful yields and limit by-products. As a result, most textiles are landfilled, contributing to microplastic pollution and material loss [1].

Fueled by a theoretical prediction and subsequent demonstrations of passive cooling [20–24], interest has recently emerged in melt-spun polyethylene (PE) yarns as a replacement for conventional textile-grade yarns. PE-based thermoplastic olefin resins (homopolymers, copolymers, and blends) can be derived from fossil fuels or biomass, require low energy and water in production, and offer excellent chemical resistance and closed-loop recyclability [25–27]. Despite this, PE use has historically been limited to technical nonwoven textiles for construction, agriculture, and protective gear (e.g., Tyvek®) and to gel-spun ultra-high molecular weight (UHMW) PE fibers for ropes, nets, and ballistic/cut protection (e.g., Dyneema®) [28,29].

Recently, key barriers to wearable PE fibers, namely non-dyeability and hydrophobicity, were overcome through spin-dyed PE fibers and fabrics with moisture-wicking capabilities superior to conventional textiles [30,31]. However, plying these PE fibers with conventional elastomer cores would still render the heterogeneous yarns non-recyclable. Meanwhile, current methods for making melt-spun PE fibers elastic convert the thermoplastic into a thermoset, eliminating its mechanical recyclability [32–34]. These limitations call for a strategy to introduce elasticity without compromising material uniformity and recyclability.

In nature, such multifunctionality often arises through structural organization rather than changes in chemical composition. Plant fibers built from cellulose derive their strength, elasticity, and moisture transport properties from how the cellulose is arranged and aligned, so different plants, or even different parts of one plant, exhibit markedly different mechanical properties despite identical chemistry.

Here, we show that, by melt-spinning chemically identical PE-based resins and their blends under different conditions, we similarly engineer the polymer's internal structure to achieve a range of mechanical properties. This eliminates the need to ply chemically dissimilar fibers to obtain the desired elastic yarn performance.

To explore the tensile properties achievable with PE fibers, a library of fibers and yarns was melt-spun from several PE resins, including olefin block copolymer (OBC), low-density PE (LDPE), linear low-density PE (LLDPE), medium-density PE (MDPE), and high-density PE (HDPE) (**Supplementary Notes 1-2**). PE chains consist of a carbon backbone $(C_2H_4)_n$, with two hydrogens per carbon. LDPE is highly branched, giving high flexibility and low strength. HDPE consists of long, linear chains with minimal branching, leading to higher crystallinity, stiffness, and strength (**Figure 1a**) [35–37]. LLDPE and MDPE are intermediate between LDPE and HDPE, reflecting differences in branching. OBCs alternate linear PE hard blocks with octene-rich soft blocks (**Figure 1a**). The hard segments provide structural rigidity, while octene comonomers in the soft blocks suppress crystallization and impart elasticity [38–41]. Despite these differences, attenuated total reflectance Fourier-transform infrared (ATR-FTIR) spectroscopy confirms that all resins used in this study retain the characteristic PE backbone (**Supplementary Note 3**). All forms of PE are nontoxic, chemically stable, and stain-resistant [26,31,42,43]; apart from high-viscosity UHMWPE [28], they can be melt-spun into continuous fibers using standard extrusion and spinning. PE resins can be synthesized from fossil-derived or sugarcane-based feedstocks, both included in this study (**Supplementary Table S1**) [31,44].

Melt-spinning and drawing of thermoplastic fibers or yarns involve the following steps [25,42]: (i) the polymer is heated above its melting point and extruded through fine holes in a spinneret to form continuous filaments; (ii) the emerging molten strands are rapidly cooled and solidified by air quenching to form as-spun fibers; and (iii) the fibers are subsequently stretched under controlled temperature and tension (drawn) to align the polymer chains along the fiber axis (**Figure 1b**). The drawing process increases chain orientation and crystallinity, yielding fibers with enhanced tensile strength, stiffness, and dimensional stability suitable for textile formation [45–51].

PE fibers and yarns were fabricated from various PE resins using both laboratory-scale extruders (DACA and Xplore) and an industrial melt-spinning system (Hills, LBS 300). The tensile property map in **Figure 1d** was constructed from room-temperature tensile testing of these fibers and yarns (**Figures 2a-b**), together with curated literature data for conventional textile materials [31,52–54]. The tested samples comprised DACA-spun PE monofilaments (~34-75 µm; **Supplementary Table S2a**), larger-diameter Xplore-spun PE monofilaments (~200 µm; **Supplementary Table S2b**), and industrially spun multi-filament PE yarns (**Supplementary Table S3**). Fiber morphology and diameters were characterized by SEM and optical microscopy (**Figure 1c, Supplementary Note 6**).

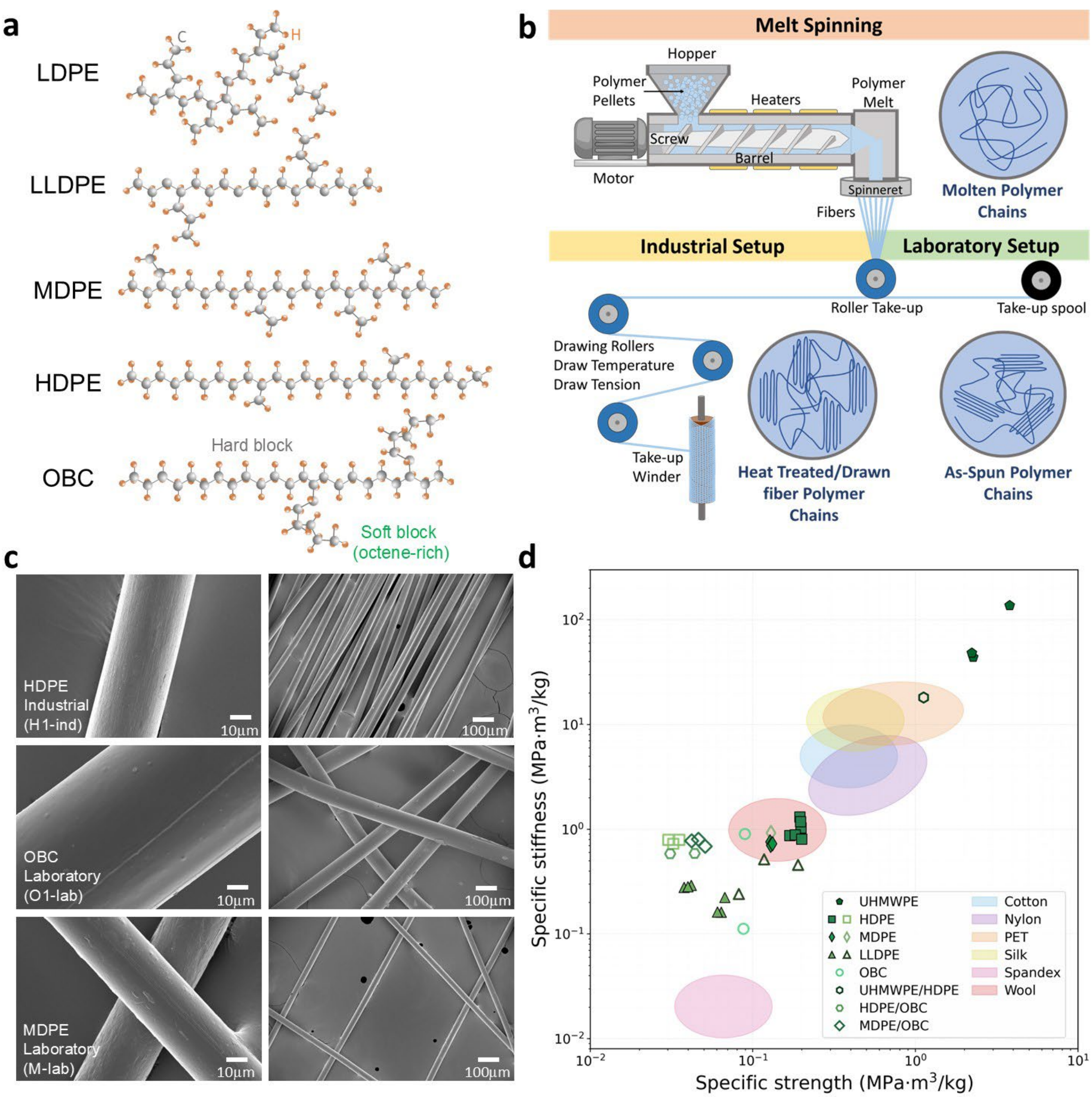

**Figure 1. A summary of molecular composition, spinning conditions, and tensile properties of PE-based filaments and yarns.** (a) Molecular structures of different PE-based resins used for fiber spinning. (b) A schematic of the fiber fabrication process and key processing parameters tuned under laboratory and industrial conditions. (c) Select SEM images of fibers produced under different conditions: HDPE yarns fabricated on the industrial spinning and drawing lines (H1-ind) and MDPE (M-lab) and OBC (O1-lab) fibers produced under laboratory conditions (**Supplementary Notes 4-6**). (d) Strength-stiffness diagram showing the tensile properties of PE-based fibers (green symbols) compared to literature data ranges of other natural and synthetic fibers (colored regions) [52,53]. Open and filled symbols represent laboratory-spun and industrially produced fibers, respectively (**Tables S2-S4**).

PE-based fibers alone span a remarkably broad mechanical space (**Figure 1d**), from the lower-left quadrant typical of single-network elastomers to the upper-right quadrant of high-performance stiff fibers. For instance, lab-spun OBC fibers (O1-lab) exhibit a Young's modulus of 0.098 GPa, being soft and highly compliant yet stiffer than commercial spandex fibers (~0.024 GPa) [10,33].

Meanwhile, industrially produced fossil- and bio-derived HDPE fibers (H1-ind, H2-ind) reach 0.97 ± 0.20 GPa, on par with wool [52]. **Figure 1d** also includes gel-spun UHMWPE fibers [28], which occupy the top-right corner with extreme stiffness and strength, and melt-spun HDPE-UHMWPE fibers [31] which can match high-tenacity PET (~18 GPa) and exceed nylon-6 (~3-4 GPa) [33,55]. These results challenge the view that separate polymer chemistries are required for elastomeric and high-strength textile performance. Note that laboratory melt-spinning systems do not reach the draw ratios of industrial lines, producing fibers with lower stiffness and tenacity.

**Figure 2** summarizes the tensile responses underlying the property map in **Figure 1d**, with Young's modulus extracted as detailed in **Supplementary Note 7**. As expected, yarns made from higher-density MDPE and HDPE resins processed under industrial conditions exhibit higher Young's moduli and tenacity values (**Figure 2a-b**). H1-ind (fossil-derived) and H2-ind (bio-derived) yarns exhibit the highest tensile strength of ~175 MPa, making them the best candidates for the sheath yarn.

Laboratory-fabricated monofilaments exhibited greater extensibility due to reduced chain alignment, allowing increased molecular rearrangement during deformation. However, high extensibility or low stiffness alone do not indicate suitability as an elastic core material, as repeated deformation performance depends on elastic recovery. Therefore, cyclic tensile tests were performed to evaluate recoverability and identify candidates capable of replacing spandex in PE-based mono-material yarns (**Figures 2c-d; Supplementary Note 7**).

O1-lab monofilament exhibited elastic recovery exceeding 90% at moderate strains up to ~140%, and remained highly extensible over several hundred percent strain, with elastic behavior comparable to that of commercial spandex fibers (**Figures 2c-d, Supplementary Figure S6**).

In contrast, conventional textile fibers such as nylon, silk, and viscose rayon typically fail at substantially lower strains (~20–25%) with poor recovery even under small deformations [52,56]. This makes O1-lab a rare combination of high extensibility and recoverability among PE-based fibers, highlighting its potential to substitute spandex as the elastic core in mono-material textile systems. Notably, the high-strength HDPE fiber H1-ind also retained measurable elastic recovery at low strains, though well below O1-lab, supporting its role as a load-bearing sheath that can accommodate some reversible deformation (**Figure 2c)**.

To link these mechanical differences to structure, the crystallinity, thermal behavior, and chain orientation of processed fibers were characterized by differential scanning calorimetry (DSC) and wide-angle X-ray scattering (WAXS) (**Supplementary Notes 9-10**), with DSC performed on as-received samples to preserve processing-induced structure. At the same time, molecular weight distributions (MWDs) of the parent resins were evaluated separately in **Supplementary Note 8**.

Fibers spun from higher-density resins generally exhibited both higher crystallinities and elevated melting points (**Figure 3a**). Industrial HDPE fiber H1-ind showed the highest crystallinity (~65%) and melting point (~130 °C), whereas elastic OBC fiber O1-lab exhibited the lowest crystallinity (~14%) and a melting temperature around 122 °C, consistent with previous reports on PE fibers fabricated by either melt-spinning or electrospinning [31,56–58].

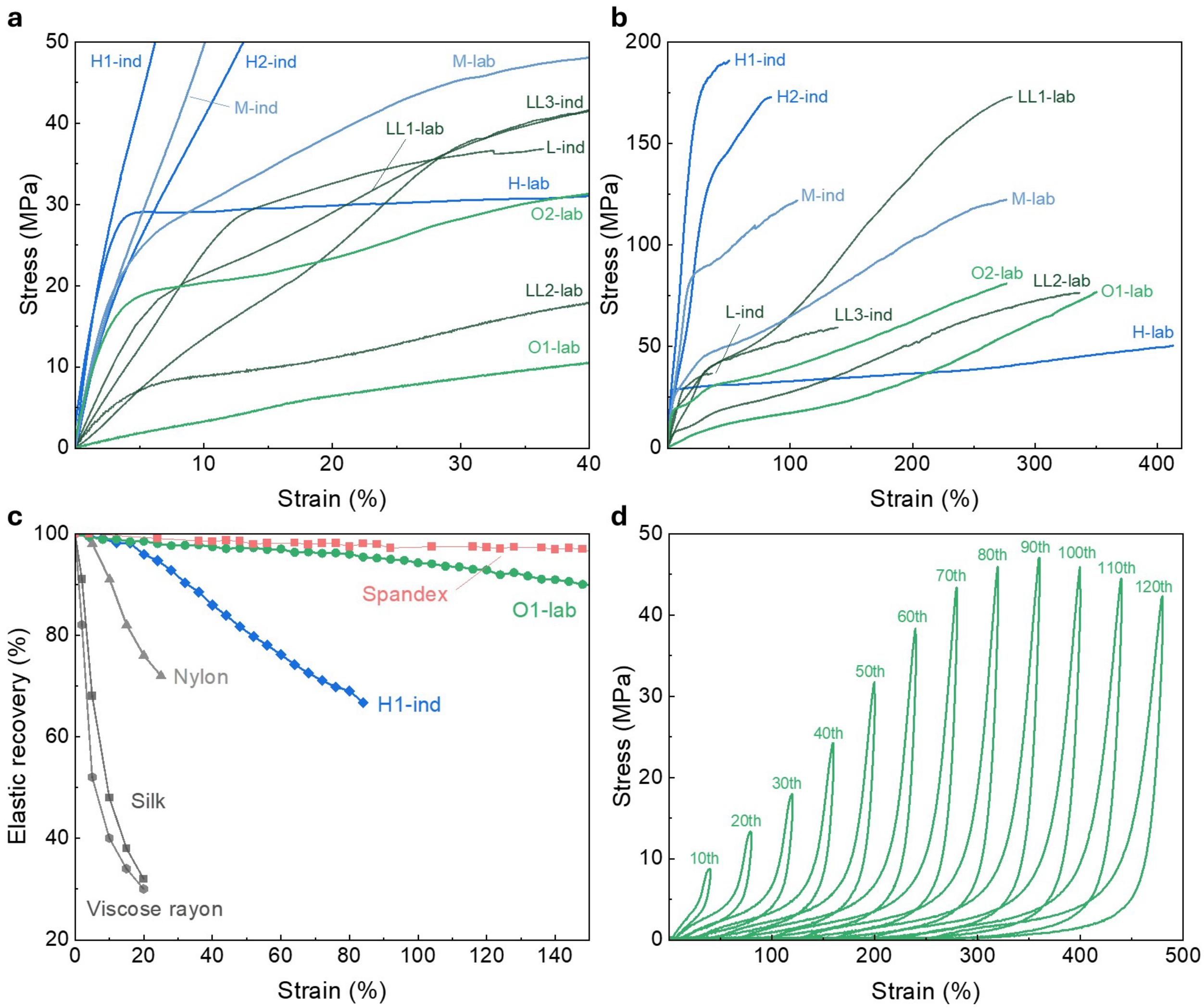

**Figure 2. Tensile strength and elastic recovery of PE-based fibers fabricated under industrial and laboratory conditions.** (a) Stress-strain curves of select PE fibers under small deformations (0-40% strain). (b) Full stress-strain curves. (c) Elastic recovery as a function of strain for O1-lab, H1-ind, and Spandex fibers compared to commercial textile fibers (nylon, silk, viscose) in the small strain range [52]. (d) Elastic recovery of O1-lab fibers over an extended strain range (up to 500%), obtained from cyclic stress-strain responses under progressively increasing strain amplitudes (1 mm increments), with representative cycles (every 10th cycle) shown **(Supplementary Figure S6)**.

The 2D WAXS patterns underlying the crystallinity and Herman's orientation factor (HOF) in **Figure 3b** are shown in **Figure 3c** for three selected fibers. Industrially processed H1-ind and M-ind fibers showed sharp equatorial arcs from the (110) and (200) reflections of the orthorhombic PE phase, consistent with higher crystallinity and partial lamellar alignment along the fiber axis (**Figure 3b**). In contrast, the OBC fiber showed low-intensity arcs and an amorphous halo, indicating suppressed crystallinity and reduced orientation.

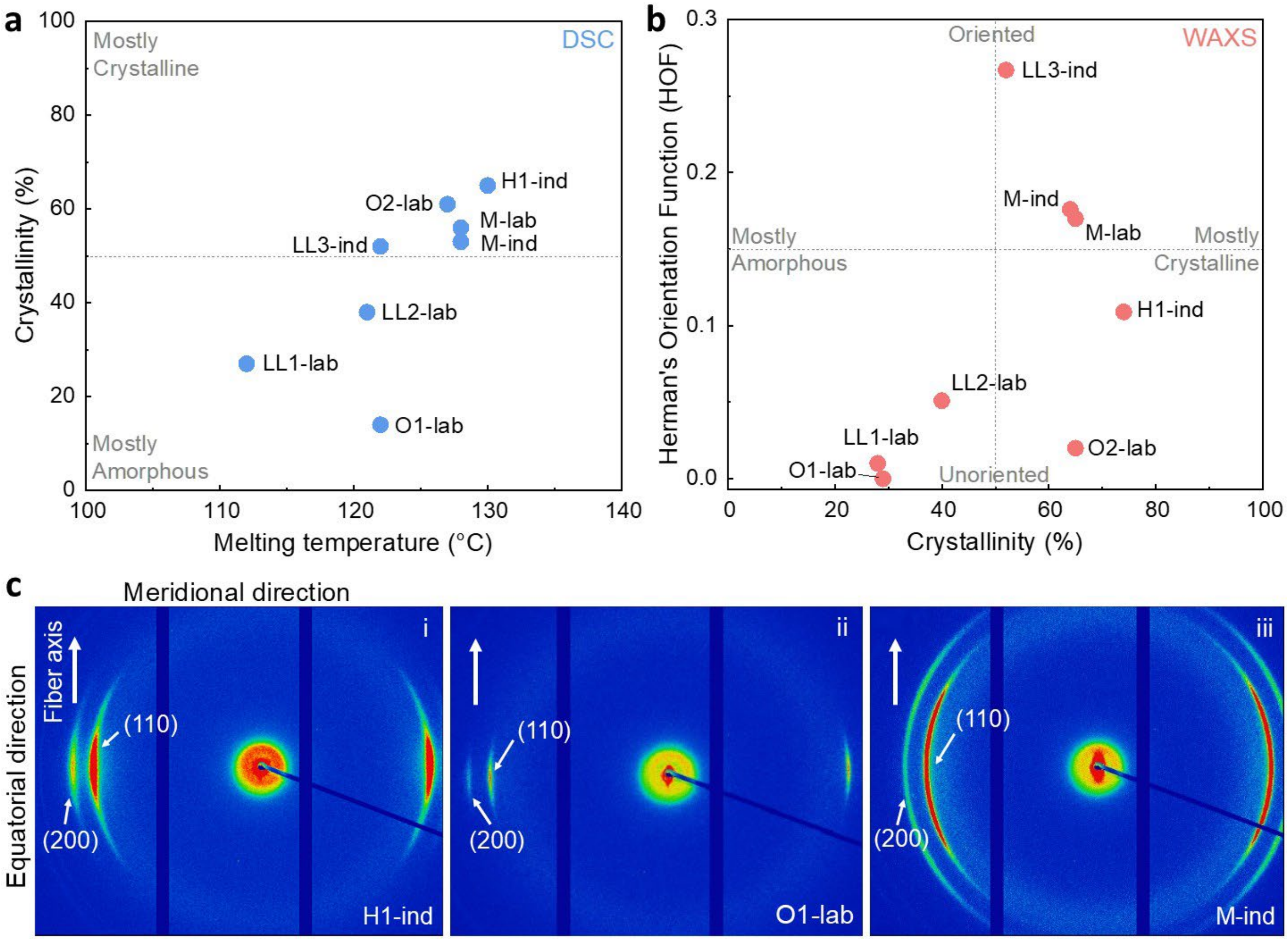

**Figure 3. Crystallinity and molecular orientation of PE-based fibers.** (a) Degree of crystallinity and melting temperature of PE fibers measured by DSC. (b) Herman's orientation function calculated from WAXS spectra and plotted against the fiber crystallinity. (c) 2D WAXS diffraction patterns for selected fibers, including H1-ind (c-i), O1-lab (c-ii), and M-ind (c-iii).

Although MWDs may influence crystallization, comparisons among the LLDPE grades suggest that comonomer chemistry and chain architecture also shape mechanical properties beyond crystallinity alone (**Supplementary Table S7**). Across the broader set of fibers, however, the largest mechanical differences remain most closely tied to the crystallinity and orientation of processed fibers, highlighting the role of processing-induced structural control in balancing load-bearing capability and extensibility in the PE-based yarns.

These structure-property relationships enabled rational selection of O1-lab fibers as elastic cores and H1-ind fibers as high-strength sheaths for recyclable PE-based stretchable yarns. The demonstrated variability of mechanical properties achievable in melt-spun PE fibers enables construction of hierarchical stretchable yarns that can replace conventional PET-spandex yarns while maintaining full thermo-mechanical recyclability.

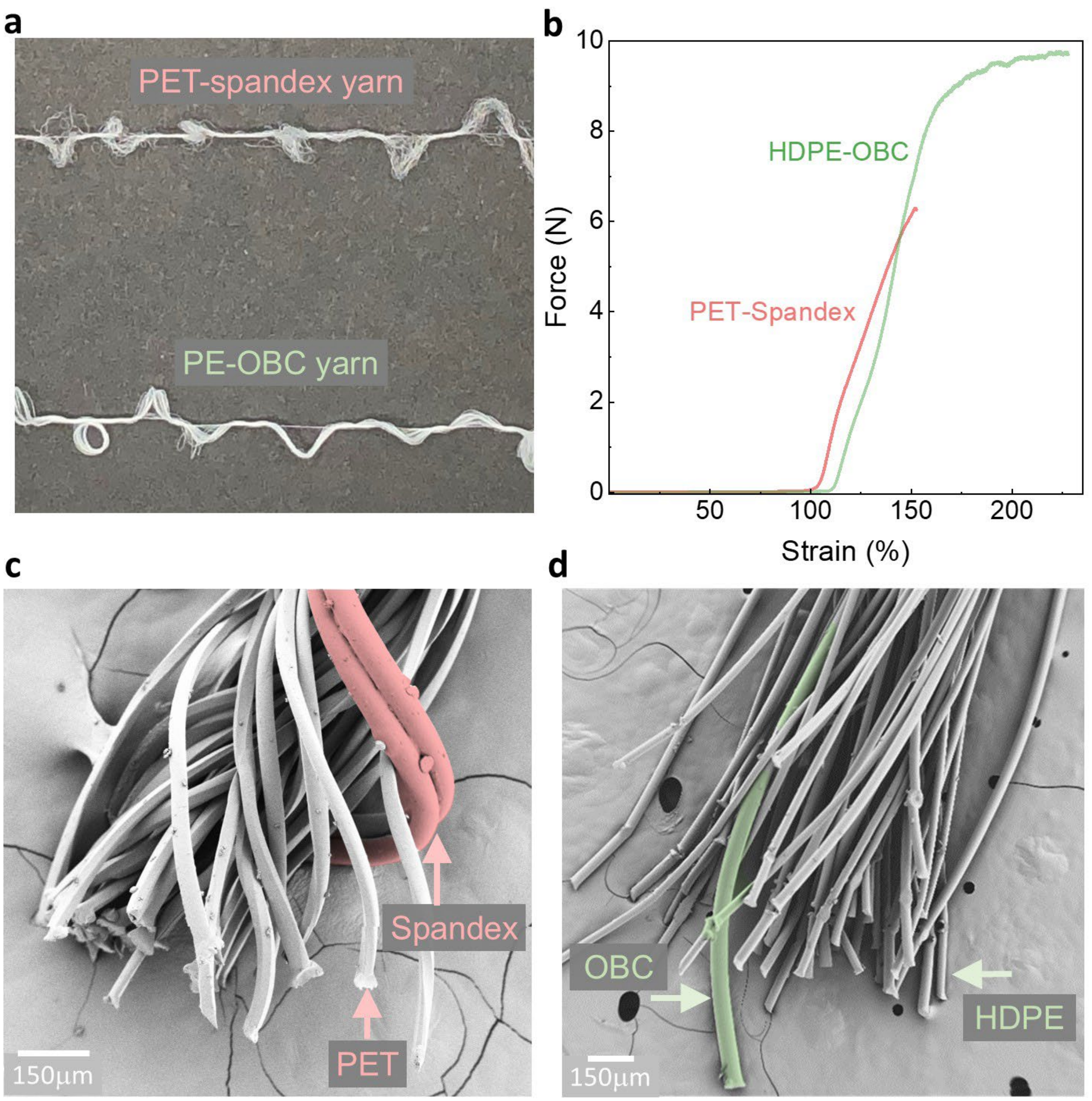

**Figure 4. Comparison between a conventional elastic yarn and the newly developed recyclable HDPE-OBC stretchable yarn.** (a) Optical images of the commercial stretchable hetero-material PET-spandex yarn (top) and the HDPE-OBC recyclable yarn (bottom). (b) Load-strain curves of the HDPE-OBC yarn (green) and the PET-spandex yarn (red)**.** (c, d) SEM images of the PET-spandex and HDPE-OBC yarns, respectively.

Guided by these findings, **Figure 4** demonstrates a recyclable PE-based stretchable yarn. An O1-lab core (74 µm, 34 denier) was co-twisted with a 45-filament, 305-denier H1-ind sheath yarn to form the core-sheath structure shown in **Figure 4a** (**Supplementary Note 13**). For comparison, a commercial elastic yarn consisting of a 34-filament PET sheath wrapped around a 20-denier dog-bone-shaped spandex core was used as a reference (**Figure 4a; Supplementary Figure S11**). In both systems, the loose loops of the sheath yarns allow their corresponding elastic cores to expand and contract. The high-tenacity sheath fibers protect the weaker core filament from breaking under tension by limiting its extension and bearing the load once the loops are fully stretched.

The tensile response of the helical yarns (**Figure 4b**) reflects sequential deformation of the stretchable core and load-bearing sheath. The initial region corresponded to stretching of the elastic cores to approximately 100% of their original length, along with the flattening of the sheath loops, whereas the steeper slope at higher strain levels corresponded to load transfer to the sheath yarns. Upon unloading, recoil of the core filament restores the helical structure by pulling the sheath loops back into position (**Supplementary Video 1**).

The HDPE-OBC yarn not only matched the elastic response of the PET-spandex yarn but also achieved a higher maximum load (**Figure 4b**). Scanning electron microscopy further revealed comparable core-sheath architectures in both systems (**Figures 4c, d**). These results indicate that recyclable PE-based yarns can be engineered to meet or exceed the mechanical performance requirements of commercial elastic yarns, although this improvement came with an increased linear density (from 170 to 339 total denier in this case).

Although HDPE and OBC possess chemically similar hydrocarbon backbones, suggesting compatibility for thermo-mechanical recycling (**Supplementary Note 3**), they differ substantially in branching architecture and mechanical behavior. To evaluate compatibility, monofilaments were melt-spun from HDPE (H-lab), OBC (O1-lab), and a 9:1 v/v HDPE-OBC blend (H-O1) on a laboratory extruder (Xplore); this ratio approximates the core-to-sheath proportion of the PE-based stretchable yarn. ATR-FTIR, DSC, and tensile testing then evaluated blend compatibility through chemical similarity, thermal behavior, and mechanical performance.

ATR-FTIR spectra of the blend (H-O1) showed characteristic hydrocarbon peaks of the parent PE resins with no new functional groups, indicating retention of the PE backbone after blending (**Supplementary Note 3, Figure 5e**). DSC showed sharp melting and crystallization peaks near those of the H-lab fiber, indicating that OBC does not severely hinder crystallization, and provided no thermal evidence of phase separation (**Figure 5a**).

The robust performance of these multi-molecular-weight blends is underscored by the role of long chains during melt-spinning. In flow-induced crystallization, longer chains initiate a highly oriented shish-kebab crystalline morphology [59], where threadlike cores (shish) are surrounded by plate-like lamellae, and nearby shorter chains join the oriented crystalline structure [38,46]. Similar reinforcement has been reported for optimized HDPE/UHMWPE/OBC blends processed by melt-spinning, injection, and compression molding [26,60-64]. OBC resins at small fractions (4-8%) have also served as compatibilizers for recycled polypropylene/HDPE or polypropylene/LLDPE blends [65].

The blended H–O1 fibers exhibited only a modest reduction in Young’s modulus relative to H-lab fibers while maintaining tensile strength in the same range (**Figure 5b; Supplementary Table S5**). This suggests OBC does not substantially compromise mechanical performance despite increased elasticity, consistent with partial co-crystallization between OBC hard segments and HDPE domains, while amorphous regions may dissipate energy and delay fracture. Together, the chemical, thermal, and mechanical results indicate good HDPE-OBC compatibility in the as-spun blend with no evidence of strong phase separation.

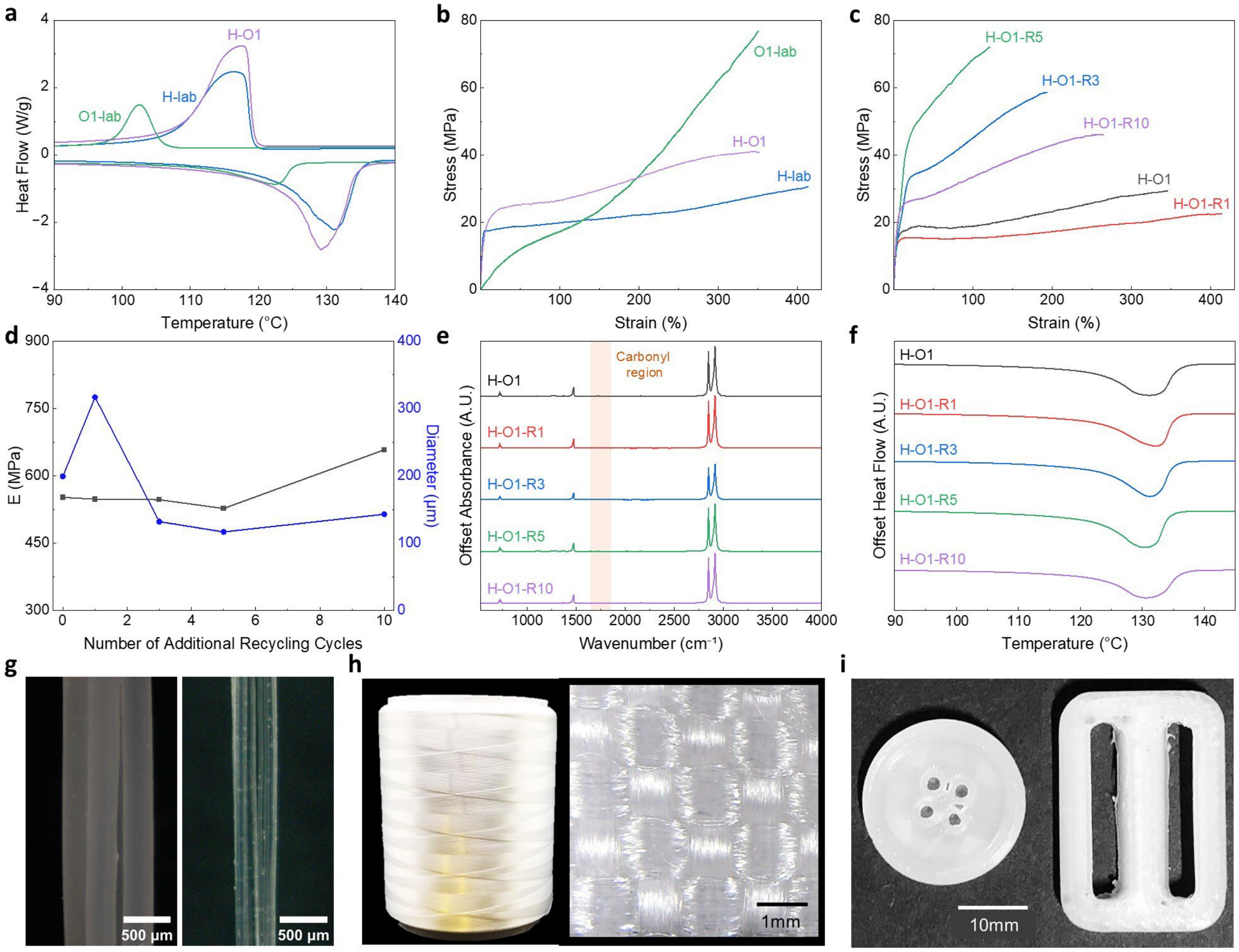

**Figure 5. Evaluation of blend compatibility and thermo-mechanical recyclability of HDPE/OBC blends.** H-O1 denotes the as-spun HDPE/OBC blend, while H-O1-Rx denotes the same blend after x additional thermo-mechanical recycling cycles. (a) DSC thermograms of HDPE, OBC, and their HDPE/OBC blend (H-O1, 9:1 v/v). (b) Tensile stress-strain curves of fibers melt-spun from HDPE, OBC, and their 9:1 HDPE/OBC blend under laboratory conditions. (c) Tensile stress-strain curves of H-O1, R1, R3, R5, and R10. (d) Young's modulus of H-O1 fibers as a function of the number of additional thermo-mechanical recycling cycles. (e) ATR-FTIR spectra and (f) DSC thermograms of H-O1, R1, R3, R5, and R10. (g) Optical microscopy images of recycled H-O1 fibers, showing representative morphologies of H-O1-R1 (left) and H-O1-R10 (right). (h) Left: Industrial-scale bobbin (0.5 kg) of melt-spun M-O1 (1:1 v/v) blended yarn. Right: A fabric sample hand-woven in a plain weave from the recycled M-O1 yarn. (i) Wearable accessories fabricated from PE-based blends via filament melt-spinning and fused filament fabrication (FFF).

Although compatibility between HDPE and OBC blends suggests potential for thermo-mechanical recycling, retention of functionality after repeated processing remains a critical requirement for practical textile applications. Repeated melt processing can induce oxidation, chain scission, and changes in crystallization behavior, which may ultimately degrade mechanical performance and limit recyclability [66,67].

To assess robustness under realistic recycling, the as-spun H-O1 blend underwent repeated melt-reprocessing and respinning cycles (**Supplementary Note 11**); samples after x additional cycles are denoted H-O1-Rx. Tensile testing, optical imaging, ATR-FTIR spectroscopy, and DSC were performed after 1, 3, 5, and 10 cycles (**Supplementary Table S10**).

Most importantly, from a practical standpoint, tensile testing showed that the recycled fibers retained useful mechanical performance, with Young's modulus remaining within the same range across cycles (**Figure 5c-d**). Although ultimate tensile strength and elongation at break varied, trading off strength against extensibility, overall toughness remains comparable with no systematic degradation trend (**Figure 5c**). Optical imaging showed smooth fibers with no major defects after repeated reprocessing (**Figure S10, Figure 5g**).

ATR-FTIR and DSC further probed chemical and thermal stability. ATR-FTIR showed no measurable carbonyl increase (1650-1850 $cm^{-1}$) [68], indicating no significant oxidative degradation (**Figure 5e**), and DSC showed a single dominant melting transition, with only minor peak-position shifts and no peak splitting (**Figure 5f**).

Together, these results indicate that the H-O1 blend tolerates repeated melt reprocessing without measurable degradation, retaining its structure and properties through at least ten recycling cycles. For practical textile recycling, however, materials must also be processable on manufacturing routes. We therefore showed scalability potential by demonstrating that blends of PE homopolymers and OBC copolymers can be processed on industrial fiber-spinning equipment. **Figure 5h** shows a bobbin of the M-O1 yarn (1:1 v/v MDPE/OBC) and a small hand-woven swatch made from it (**Supplementary Note 14**), demonstrating the feasibility of converting recyclable PE blends into textile-grade yarns.

Finally, beyond fiber spinning, our previous work demonstrated that optimally balanced blends of PE resins can be used for melt-spinning of filaments and prototyping of high-quality 3D-printed structures via fused filament fabrication (FFF) (**Figure 5i**) [69]. Garment accessories such as buckles, zippers, buttons, and hooks could be fabricated from MDPE- or HDPE-based blends with OBC and recycled together with the rest of the garment without disassembly through injection molding, compression molding, or FFF 3D printing [60,70].

These results suggest that all-PE garments can be mechanically recycled and reprocessed without separating the elastic core from the functional sheath. The recycled blends can then be re-adjusted with virgin or recycled PE resins to optimize performance for specific applications. PE-based products such as milk bottles are already accepted in curbside recycling [71,72], suggesting that a similar framework could be adapted for textiles, although high recycling rates will require better collection, sorting, labeling, dedicated streams, and consumer education. Even with current waste-management limits, an all-PE system is far more recyclable than inherently non-recyclable multi-polymer textiles, and recovered hydrocarbons can be reused as feedstocks for new high-molecular-weight materials or low-molecular-weight waxes, supporting a circular economy [1,8,73,74].

Overall, these findings demonstrate that stretchability and high-strength textile functions can be integrated within a single PE family platform through control of fiber structure rather than changes in polymer class. Combining an elastic OBC core with a high-strength PE sheath, we developed all-PE stretchable yarns that match or exceed conventional PET-spandex yarns while remaining thermo-mechanically recyclable. The demonstrated blend compatibility, reprocessability, and industrial-scale spinnability support this approach for mono-material circular textiles.

## ACKNOWLEDGMENTS

This work was supported by the DEVCOM Soldier Center through the US Army Research Office (W911NF-13-D-0001), The Office of Naval Research (ONR) Global via Tecnologico de Monterrey (Mexico), and the MIT Portugal Program. D.X. acknowledges the Tung Global Collaborative Fellowship. We thank Brandon Henry, Michael Lampkin, Tony DeLaHoz and Jeff Haggard (Hills Inc.) and Betty Ann Welsh (DEVCOM SC) for the help with the PE fibers and yarns fabrication, the Dow Chemical Company and Braskem for providing polymer resins, Providence Yarns for providing multi-filament HDPE yarns, and FLEX for donating commercial core-sheath elastic fibers. We are also grateful to Maren Cattonar (NERAMCO), Richard M. Osgood III (DEVCOM SC), Edward Greer (Dow), Yijian Lin (Dow), and the Fibernamics team (Portugal) for useful discussions.

## AUTHOR CONTRIBUTIONS

S.V.B. conceived the idea and supervised the research. V.K., D.X., S.-H.K., and D.J.B. fabricated fibers and yarns. S.-H.K., V.K., D.X., and D.R.F.H. performed structural, spectral, thermal, and mechanical characterization experiments and analyzed the data. K.M. performed rheological testing of polymers. S.-H.K., D.X., and D.R.F.H. edited and organized the manuscript visuals. All the authors contributed to the manuscript discussions, writing, and editing.

## COMPETING INTERESTS

A U.S. Utility Patent Application No 18/996,643 “High-performance mono-material hybrid yarns and textiles” has been filed on June 26, 2023, inventors: S.V. Boriskina and V. Korolovych.

# Supporting Information

# Polyethylene-based thermo-mechanically recyclable stretchable yarns for circular sustainable textiles

SeongHyeon Kim[1], Duo Xu[1], Volodymyr Korolovych[3], Domingo R. Flores-Hernandez[1,2], Kaniz Moriam[1], Daniel J. Braconnier[1], and Svetlana V. Boriskina[1*]

[1] Massachusetts Institute of Technology, Department of Mechanical Engineering, Cambridge, MA, 02139, USA

[2] Tecnologico de Monterrey, Escuela de Ingenieria y Ciencias, Ave. Eugenio Garza Sada 2501, Monterrey 64849, Mexico

[3] Massachusetts Institute of Technology, Institute for Soldier Nanotechnologies, Cambridge, MA, 02139, USA

## Table of Contents

* Corresponding Author E-mail: sborisk@mit.edu

## 1. Molecular formulas and typical density ranges of polymers used in stretchable yarns

**a** Polyethylene terephthalate (Polyester, PET):

Polyether-polyurea copolymer (Spandex, Lycra, Elastane):

Polyethylene (PE):

Olefin block copolymer (OBC, Infuse):

**b**

| Polyethylene type | Molecular weight (g/mol) | Density (g/cm$^3$) |
|---|---|---|
| LDPE | 30,000–50,000 | 0.87–0.925 |
| LLDPE | 40,000-90,000 | 0.91-0.925 |
| MDPE | 60,000–100,000 | 0.926–0.94 |
| HDPE | 200,000–500,000 | 0.94–0.965 |
| UHMWPE | 4–6 Million | 0.93–0.97 |

**Supplementary Figure S1. Overview of different polymers used for making fibers and yarns studied in this work.** (a) Chemical formulas of polyester (PET), spandex, polyethylene (PE), and the soft block of the olefin block co-polymer (OBC; hard block is a linear PE chain). (b) The conventional ranges of molecular weight and density of various types of PE resins.

## 2. The list of PE-based resins used for fiber spinning

**Supplementary Table S1**

| Resin type | Resin name | Manufacturer | Material origin | Label | Melt Flow Index (MFI), g/10min |
|---|---|---|---|---|---|
| OBC | Infuse 9100 | Dow | fossil | **O1** | 1.0 |
| OBC | Infuse 9500 | Dow | fossil | **O2** | 5.0 |
| LDPE | SLD 4004 | Braskem | biomass | **L** | 2.0 |
| LLDPE | Elite AT 6111 | Dow | fossil | **LL1** | 3.7 |
| LLDPE | 1648 | Dow | fossil | **LL2** | 3.5 |
| LLDPE | N/A | MiniFibers | fossil | **LL3** | N/A |
| MDPE | XUS 81841.27 | Dow | fossil | **M** | 6.0 |
| HDPE | Exp-22-BT-1405 | Dow | fossil | **H** | 1.5 |
| HDPE | N/A | Providence Yarns | fossil | **H1** | N/A |
| HDPE | SHE150 | Braskem | biomass | **H2** | 1.0 |

All MFI values were obtained from the manufacturer-provided technical datasheets, measured under the standard ASTM D1238 test conditions for polyethylene (190 °C, 2.16 kg load).

## 3. Spectroscopic confirmation of the PE backbone across resins

ATR-FTIR spectra were acquired for the neat PE-based resins (HDPE, MDPE, LDPE, LLDPE, OBC) to verify their classification as members of the polyethylene family. All spectra exhibit the characteristic absorption features of polyethylene, including $CH_2$ stretching near 2915 and 2850 $cm^{-1}$, $CH_2$ bending near 1470 $cm^{-1}$, and $CH_2$ rocking near 720 $cm^{-1}$ [1–5].

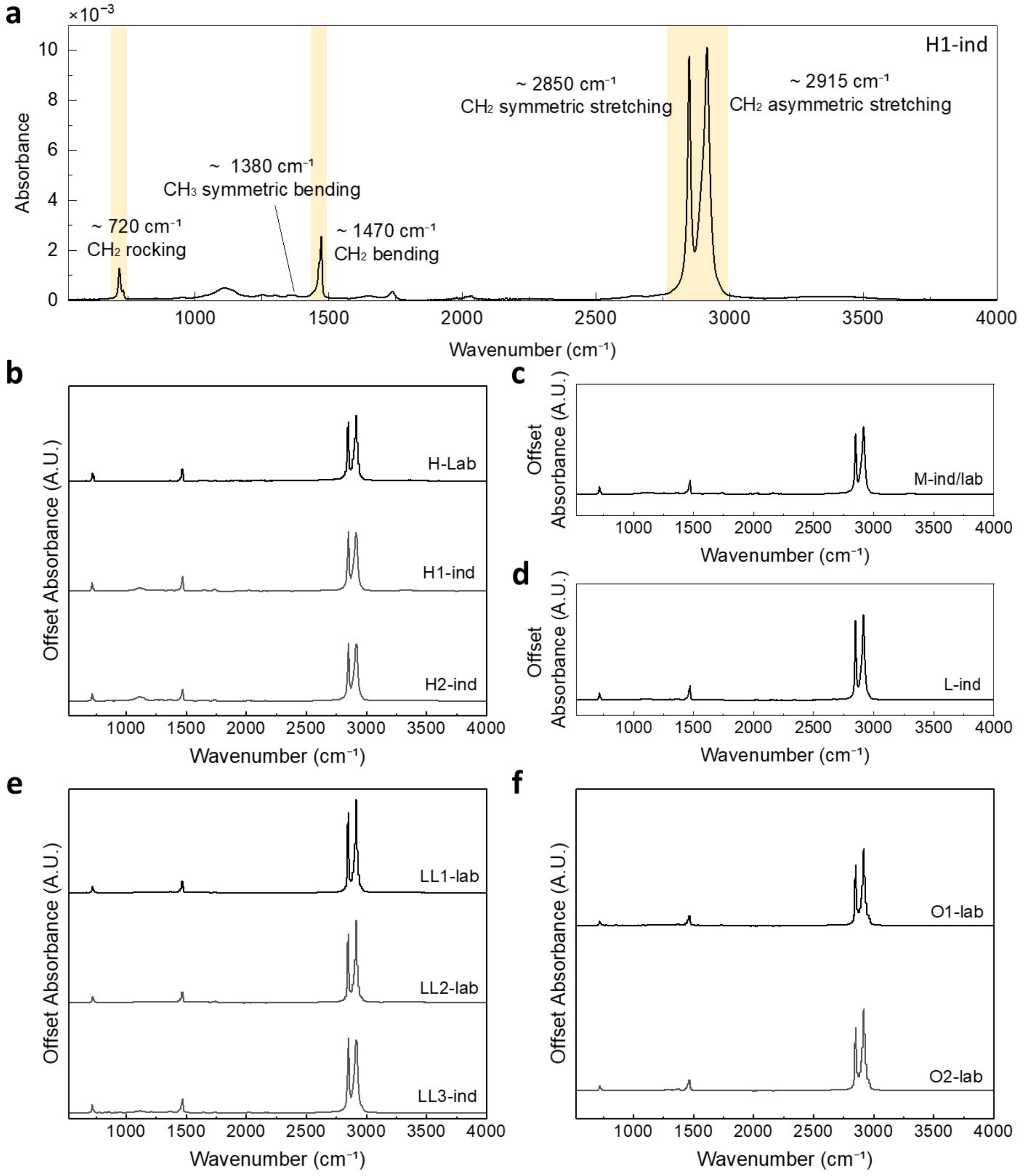

**Supplementary Figure S2. ATR-FTIR spectra of PE-based resins.** (a) Band assignments shown on the representative H1-ind spectrum [1–6]. (b) HDPE, (c) MDPE, (d) LDPE, (e) LLDPE, and (f) OBC resins. Spectra in (b–f) are area-normalized and vertically offset for clarity.

## 4. Equipment used for fiber spinning and drawing

(a)

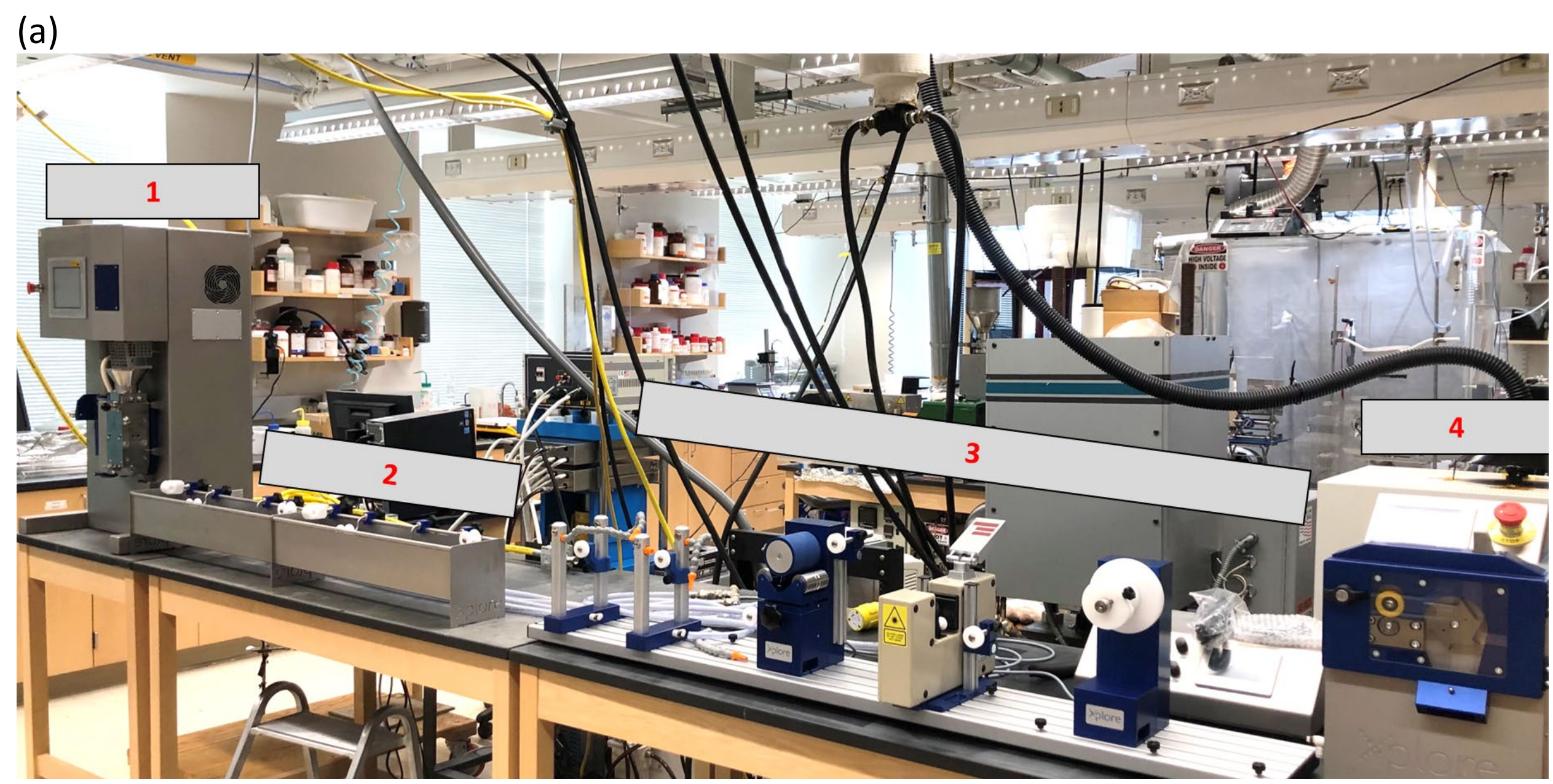

(b) (c)

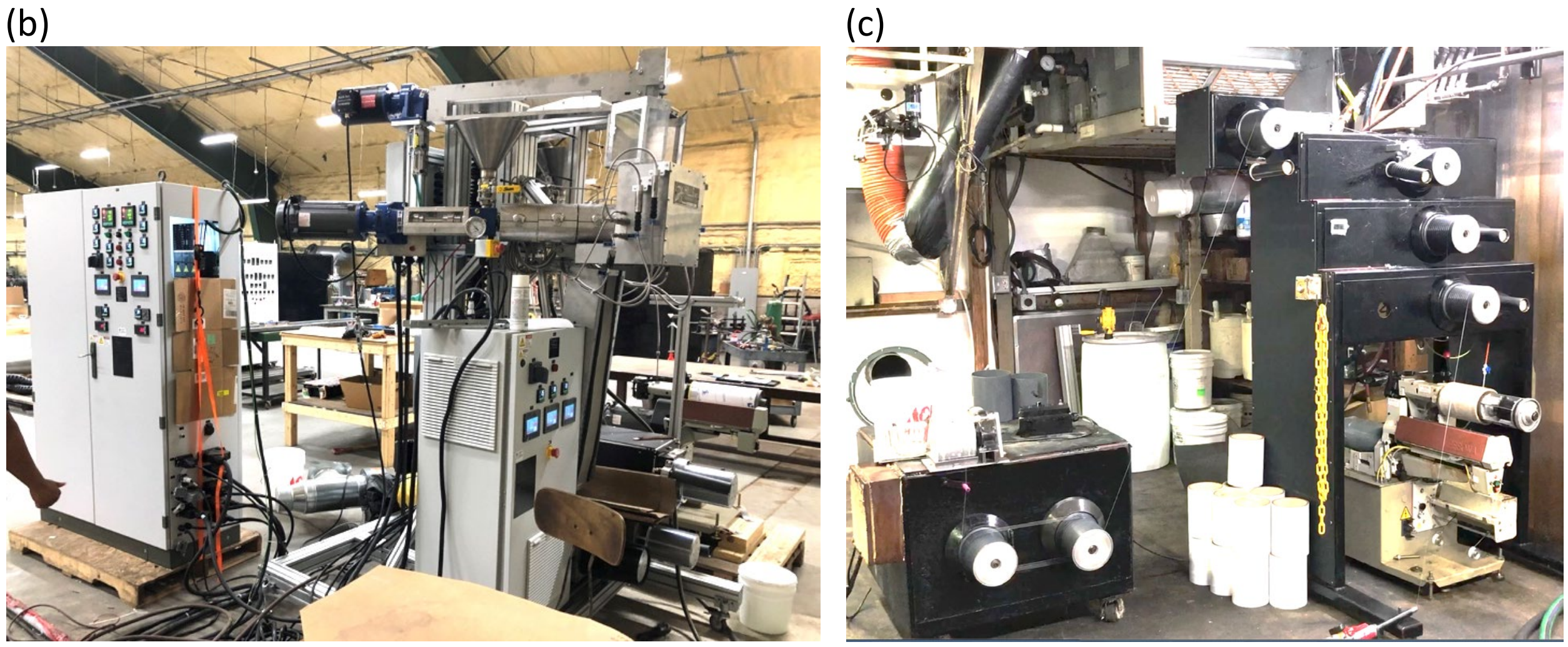

**Supplementary Figure S3. Equipment used for the fiber and yarn fabrication.** (a) A lab-scale twin-screw Xplore system including (1) micro-compounder, (2) cooling bath, (3) filament line, and (4) pelletizer. (b) An industrial multi-filament extruder (LBS 300 Hills Inc.) and a drawing system composed of denier, tension, draw, and relax rolls and a leesona winder. Winding and drawing yarns on the LBS 300 extruder was done with Lurol PP-3771 spin-finish (Goulston Technologies, Inc).

## 5. A summary of the fabricated fiber geometrical and material properties

**Supplementary Table S2.** Geometrical and material properties of the lab-spun PE monofilaments

**(a)** Textile-grade fibers spun on the DACA line with diameters comparable to commercial yarns

| Label | Material | Density, g/cm³ | Fiber diameter, µm | Production line | Processing temperature, °C |
|---|---|---|---|---|---|
| O1-lab | Infuse 9100 | 0.877 | 73.8 | DACA | 220 |
| O2-lab | Infuse 9500 | 0.91 | 34.5 | DACA | 220 |
| LL1-lab | Elite AT 6111 | 0.91 | 37.0 | DACA | 220 |
| LL2-lab | 1648 | 0.93 | 67.8 | DACA | 220 |
| M-lab | XUS 81841.27 | 0.947 | 44.7 | DACA | 220 |

**(b)** Demonstration monofilaments spun on the Xplore line for the HDPE/OBC recycling study

| Label | Material | Density, g/cm³ | Fiber diameter, µm | Production line | Processing temperature, °C |
|---|---|---|---|---|---|
| H-lab | Exp-22-BT-1405 | 0.955 | 255.9 | Xplore | 165 |
| H-O1 | Exp-22-BT-1405 + Infuse 9100 (9:1 v/v) | 0.947 | 216.6 | Xplore | 175 |

**Supplementary Table S3.** Geometrical and material properties of the industrially spun PE yarns

| Label | Material | Total denier | Density, g/cm³ | # of filaments | Measured diam, µm | Production line |
|---|---|---|---|---|---|---|
| L-ind | SLD 4004 | 1365 | 0.923 | 72 | 63.0 | Hills |
| LL3-ind[7] | LLDPE | 494 | 0.93 | 247 | 18.5 | MiniFibers |
| M-ind | XUS 81841.27 | 890 | 0.947 | 72 | 28.2 | Hills |
| H1-ind | HDPE | 305 | 0.97 | 45 | 25.1 | Providence yarns |
| H2-ind | SHE150 | 858 | 0.948 | 72 | 43.5 | Hills |
| M-O1 | XUS 81841.27 + Infuse 9100 (1:1 v/v) | 801 | 0.912 | 72 | 54.9 | Hills |

## 6. Fiber morphology and size characterization via scanning electron microscopy

Scanning electron microscopy (SEM) images were obtained using a Zeiss (model is 300 Sigma VP) microscope at a 3 kV accelerating voltage. A DESK IV Cold sputter was used to apply an 8 nm-thick gold coating to all fibers prior to imaging. Average fiber diameters reported in **Supplementary Tables S2a and S3** were quantified from SEM images using ImageJ software, as averages of at least 10 measurements per sample, whereas diameters in **Supplementary Table S2b** were obtained from optical images. **Supplementary Figure S4** shows SEM images of select PE-based fibers. All the fabricated fibers exhibited circular cross-sections.

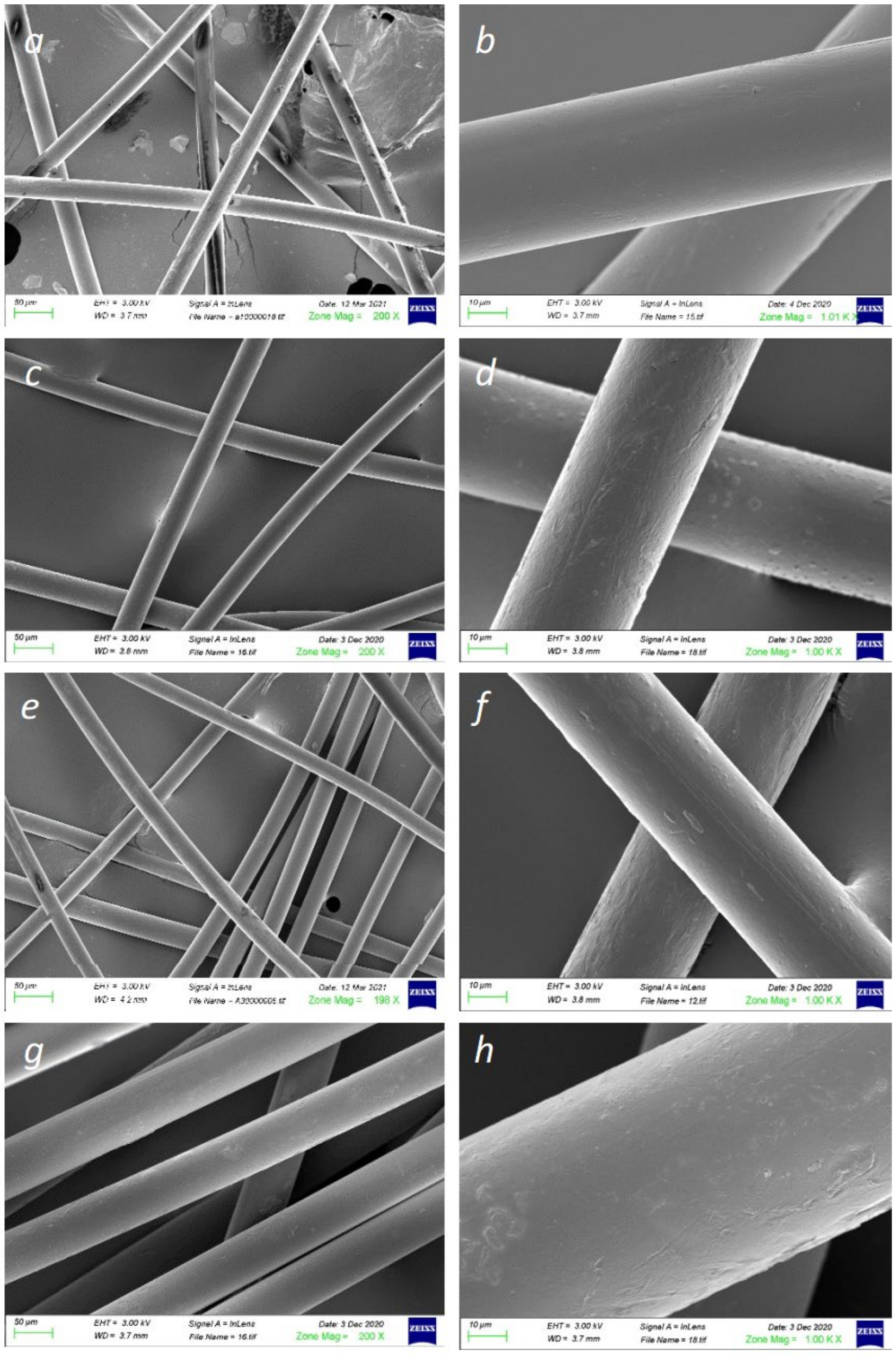

**Supplementary Figure S4.** SEM images of LL1-lab (a,b), O2-lab (c,d), M-ind (e,f) and LL2-lab (g,h) fibers at two different magnifications.

## 7. Fiber mechanical properties characterization via tensile testing

Mechanical properties of the yarns and fibers were measured using a Zwick tester (ZwickRoell Group) and an OmniTest 10 system. The gauge length was 25 mm for fibers and 54 mm for yarns. Tensile tests were conducted at a constant extension rate of 15 mm/min. All tests were conducted at room temperature. The initial portions of yarn stress-strain curves often exhibit non-linear behavior due to slack, crimp straightening, grip seating, or system compliance. To account for this variability across samples, a fixed strain threshold was not imposed; instead, Young's modulus (E) was determined using a data-driven approach. A sliding window with a fixed strain width of 0.2% was applied in the low-strain regime. Linear regression was performed in each window, and the coefficient of determination ($R^2$) was used to identify the most linear region. The slope of this region was taken as E. The starting point of this region was defined as the onset of stable elastic deformation. To further characterize stiffness over a broader range of deformations beyond the small-strain regime represented by Young's modulus in **Figure 1d**, a secant modulus was additionally calculated; the resulting diagram is shown in **Supplementary Figure S5**. Unlike Young's modulus, which primarily reflects the initial elastic stiffness, the secant modulus captures the average stiffness over larger deformation ranges that are more relevant to practical textile use. For each sample, the secant modulus was calculated over a fixed strain interval ($\Delta\varepsilon$ = 4%) starting from the identified reference point [7–10]. Local averaging was applied at the interval boundaries to reduce sensitivity to noise and discretization.

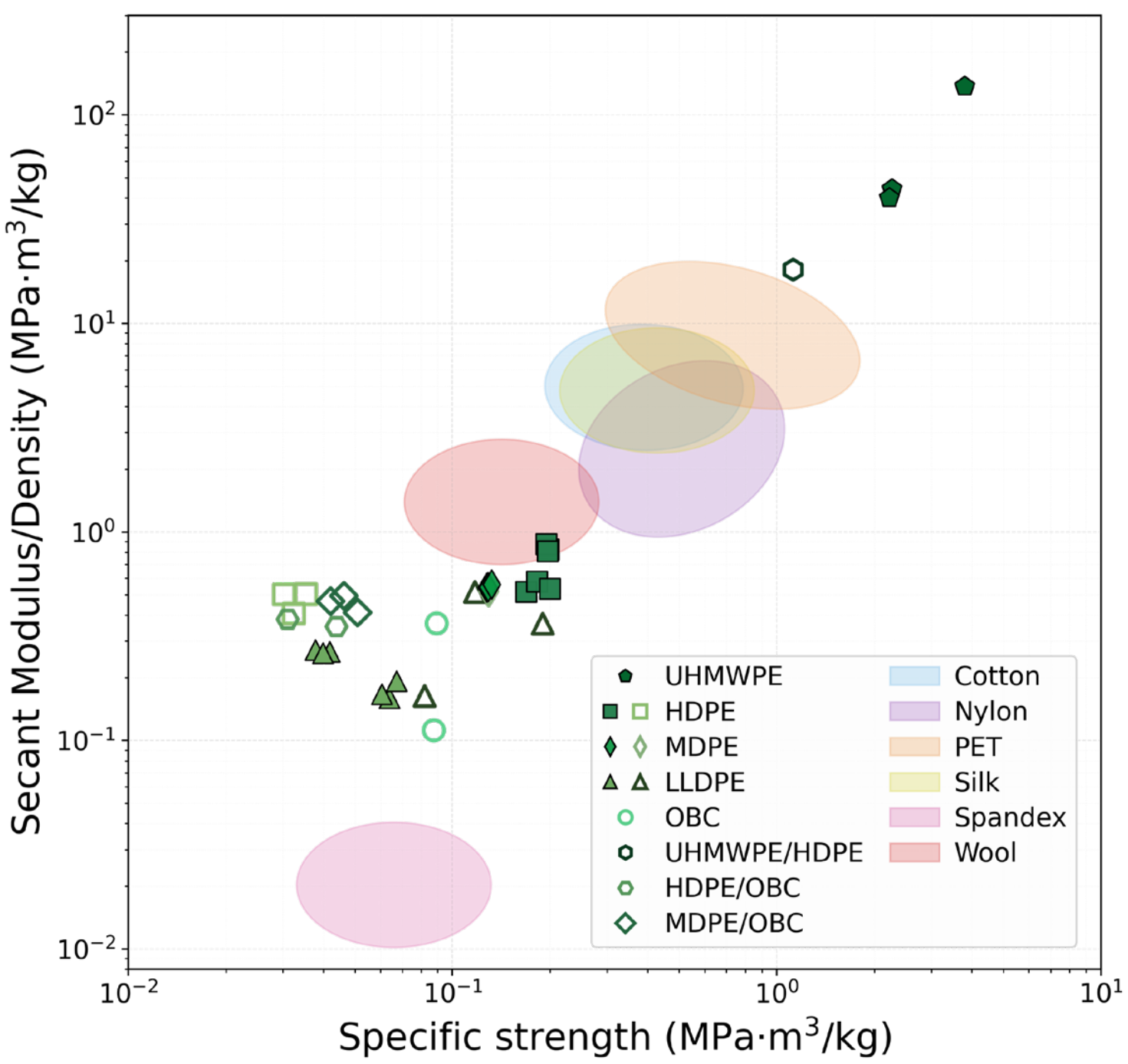

**Supplementary Figure S5.** Strength-stiffness diagram based on the secant modulus ($E_{sec}/\rho$), analogous to Fig. 1d, which is based on Young's modulus ($E/\rho$). PE-based fibers (green symbols) are compared with literature data ranges of other natural and synthetic fibers (colored regions). Open and filled symbols represent laboratory-spun and industrially produced fibers, respectively.

**Supplementary Table S4. Mechanical properties of PE fibers and yarns**

| Label | Maximum Force, N | Ultimate Tensile strength, MPa | Elongation at break, % | Young's modulus, MPa | Specific strength, MPa·m$^3$/kg | Specific stiffness, MPa·m$^3$/kg |
|---|---|---|---|---|---|---|
| O1-lab | 0.329 | 76.84 | 350 | 98.16 | 0.088 | 0.112 |
| O2-lab | 0.076 | 81.37 | 277 | 819.1 | 0.089 | 0.900 |
| LL1-lab | 0.186 | 172.82 | 281 | 414.9 | 0.190 | 0.456 |
| LL2-lab | 0.276 | 76.33 | 336 | 223.7 | 0.082 | 0.241 |
| M-lab | 0.076 | 122.54 | 276 | 872.9 | 0.129 | 0.922 |
| L-ind | 8.259 | 36.78 | 37 | 261.4 | 0.040 | 0.283 |
| LL3-ind[7] | 3.939 | 59.33 | 139 | 149.8 | 0.064 | 0.161 |
| M-ind | 5.478 | 121.81 | 106 | 666.9 | 0.129 | 0.704 |
| H1-ind | 7.075 | 190.69 | 50 | 1146 | 0.197 | 1.181 |
| H2-ind | 18.496 | 172.85 | 84 | 834.5 | 0.182 | 0.880 |

**Supplementary Table S5. Mechanical properties of H-lab and H-O1**

| Label | Maximum Force, N | Ultimate Tensile strength, MPa | Elongation at break, % | Young's modulus, MPa | Specific strength, MPa·m$^3$/kg | Specific stiffness, MPa·m$^3$/kg |
|---|---|---|---|---|---|---|
| H-lab | 7.957 | 30.94 | 411 | 693.3 | 0.032 | 0.726 |
| H-O1 | 7.659 | 41.58 | 345 | 556.3 | 0.044 | 0.587 |

**Supplementary Table S6. Mechanical properties of M-O1**

| Label | Maximum Force, N | Ultimate Tensile strength, MPa | Elongation at break, % | Young's modulus, MPa | Specific strength, MPa·m$^3$/kg | Specific stiffness, MPa·m$^3$/kg |
|---|---|---|---|---|---|---|
| M-O1 | 10.142 | 58.91 | 158 | 65.67 | 0.065 | 0.072 |

Cyclic tensile tests were performed to characterize elastic recovery of fibers. The strain amplitude was increased stepwise by 1 mm in each cycle without relaxation time between cycles. Elastic recovery was defined as the ratio of elastic extension to total extension (**Supplementary Fig. S6**) [8].

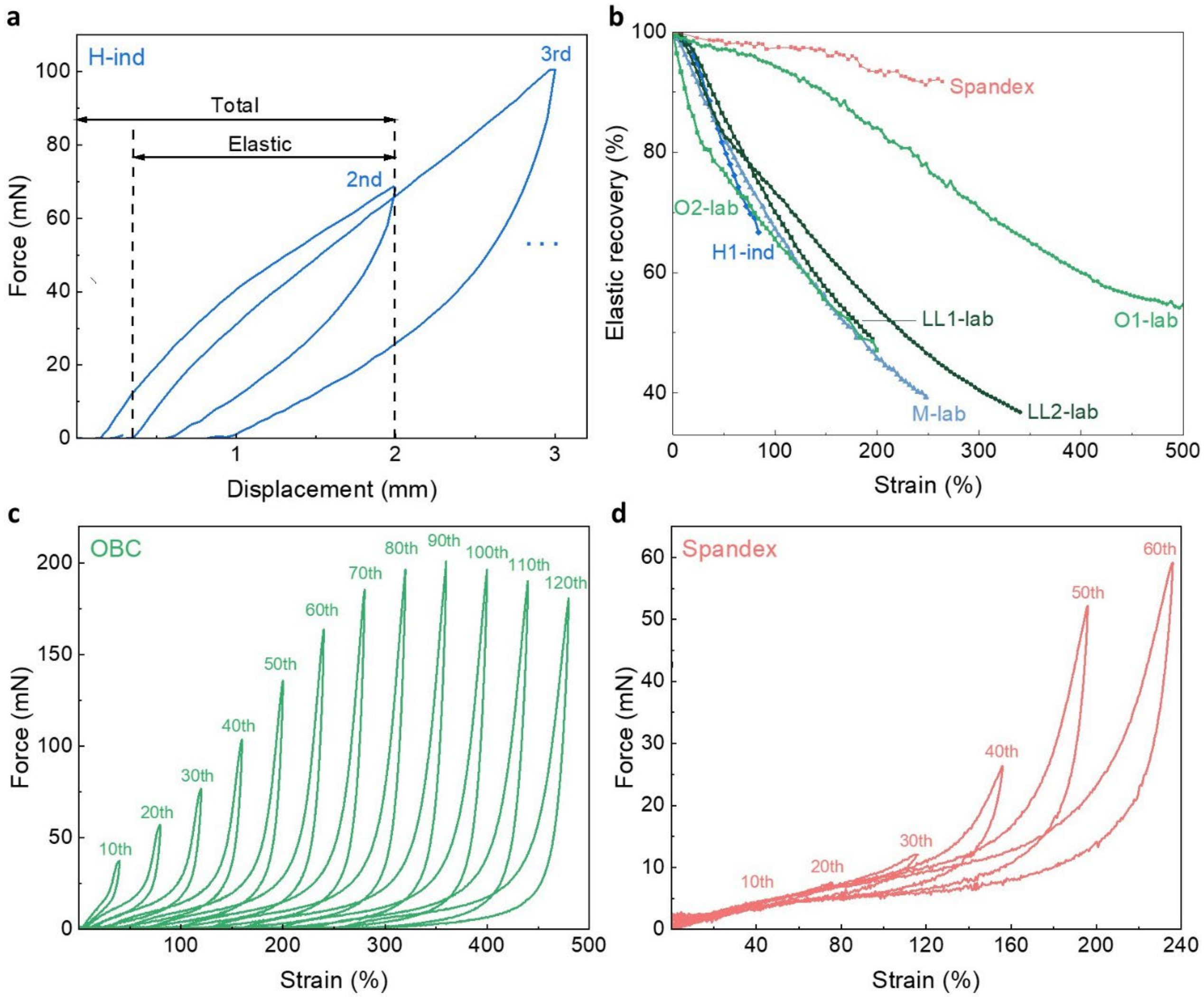

**Supplementary Figure S6. Cyclic tensile testing of fibers to probe elastic recovery.** (a) Cyclic tensile test of an H1-ind fiber at progressively increasing elongation levels, with strain amplitude increased stepwise by 1 mm, showing the 2nd and 3rd cycles and indicating the elastic and total elongation. (b) Elastic recovery as a function of strain for selected PE-based fibers. (c,d) Cyclic tensile tests at progressively increasing elongation levels, with every 10th cycle shown for an O1-lab fiber (c) and a spandex fiber (d).

## 8. Molecular weight distributions from small-amplitude oscillatory shear rheology

### Rheological Characterization

Shear rheology was measured using a Discovery Hybrid Rheometer 3 (TA Instruments, New Castle, DE) with 25 mm aluminum parallel plates (1 mm gap) and Peltier temperature control.

Small-amplitude oscillatory shear (SAOS) frequency sweeps were performed at a strain amplitude of 0.5% (within the linear viscoelastic regime), over 0.1-600 rad $s^{-1}$ for HDPE and LLDPE, and 0.1-100 rad $s^{-1}$ for OBC, at 150–210 °C. Master curves of the storage modulus $G'(\omega)$, loss modulus $G''(\omega)$, and complex viscosity were generated by time–temperature superposition (tTS, $T_{ref} = 150$ °C) in TRIOS v5.1.1.46572, and the resulting shift factors were fit to the WLF equation.

### Molecular weight distribution from rheology

The polymers studied here were compared in terms of their melt rheology and polydispersity, both estimated from the SAOS measurements described above. Since the aim was to assess how polydispersity influences yarn properties, this analysis was restricted to polymers available as multiple resin grades (HDPE, LLDPE, and OBC), so that within-polymer comparisons across grades were possible; single-grade polymers such as MDPE were excluded as no internal comparison could be made.

For each polymer, molecular weight distributions (MWDs) were estimated from its linear viscoelastic response through relaxation-spectrum analysis. The dynamic moduli measured by SAOS were fitted to the generalized Maxwell model in its discrete form to obtain the relaxation time spectrum g(τ) [11,12]. The inversion was carried out on a set of discrete τ values spanning the experimentally accessible frequency range. Non-negativity of g(τ) enforced through a non-negative least squares (NNLS) solver [13]; Tikhonov regularization at $\lambda = 5\times10^{-4}$ was applied to obtain a stable relaxation spectrum and minimize numerical artifacts in the reconstructed spectrum [14].

The relaxation time spectrum was converted to a molecular weight distribution using a power-law relationship $\tau \propto M^{a}$, with the exponent a = 3.4 reflecting reptation scaling of entangled linear chains [15]. The prefactor $k = 10^{-18.3}$ was adopted from the literature as a calibration constant representative of polyolefin-like systems [16,17]. As prefactor k is sensitive to both chemistry and temperature, the resulting MWDs should be interpreted as relative rather than absolute molecular weight distribution. A single value of (k) was applied consistently across all samples to enable direct comparison of distribution shape, breadth, and high-molecular-weight-tail contributions. The resulting molecular weight distributions, $w(log\ M)$ curves were normalized to unit area.

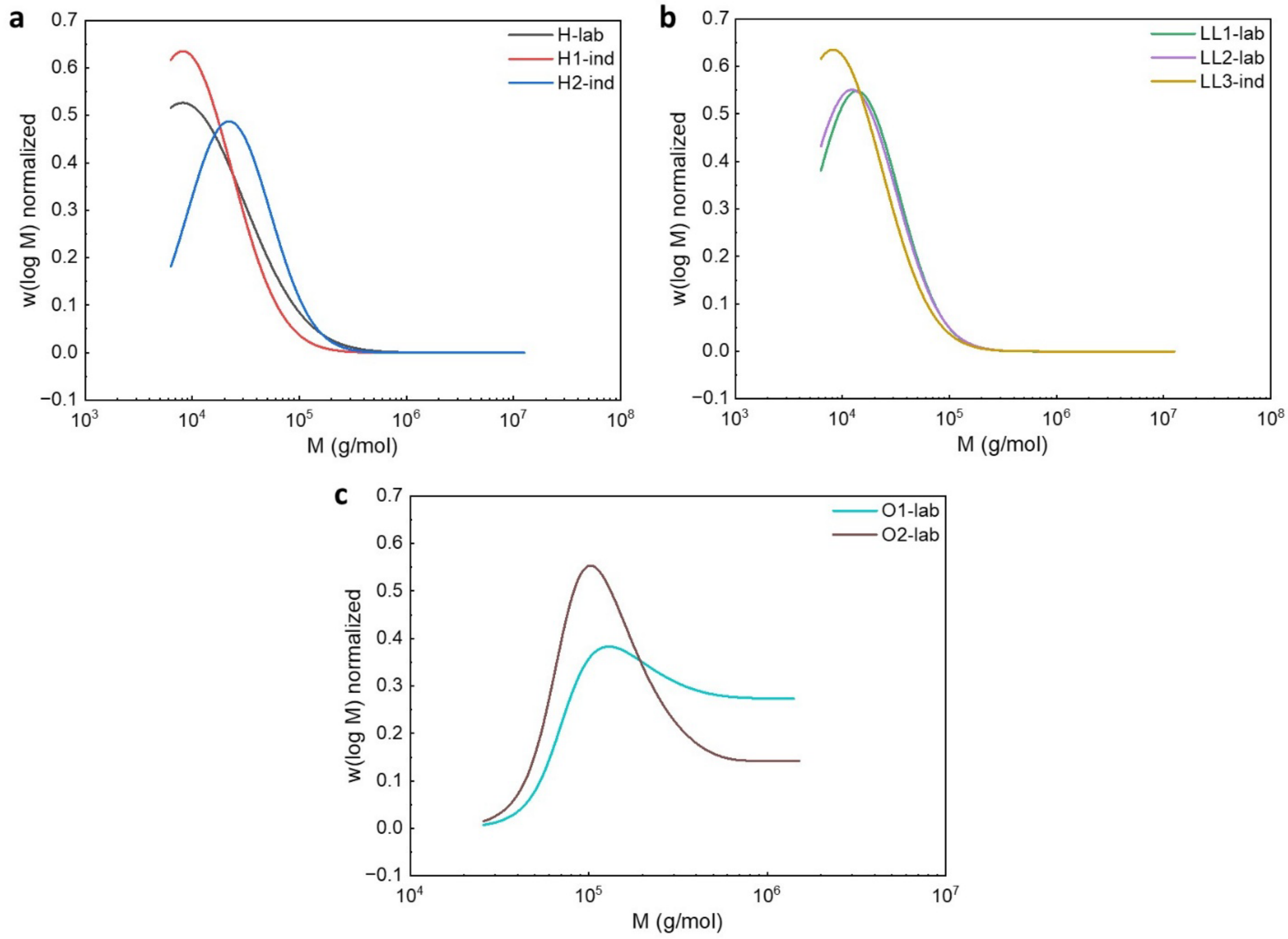

**Supplementary Figure S7. Molecular weight distributions of PE-based resins inferred from melt rheology.** Normalized w(log *M*) curves for (**a**) HDPE, (**b**) LLDPE, and (**c**) OBC resin grades, obtained from SAOS data as described above.

**Supplementary Table S7. Molecular weight averages and PDI of PE-based resins**

| Label | Resin | Mn (g/mol) | Mw (g/mol) | Mz (g/mol) | PDI |
|---|---|---|---|---|---|
| H-lab | Exp-22-BT-1405 | 29339 | 43089 | 74569 | 1.47 |
| H1-ind | HDPE | 25892 | 33448 | 50282 | 1.29 |
| H2-ind | SHE150 | 30475 | 41543 | 61254 | 1.36 |
| LL1-lab | Elite AT 6111 | 26592 | 33456 | 45991 | 1.26 |
| LL2-lab | 1648 | 26703 | 34223 | 48474 | 1.28 |
| LL3-ind[7] | LLDPE | 23003 | 26692 | 33220 | 1.16 |
| O1-lab | Infuse 9100 | 175353.5 | 390902.1 | 705869 | 2.23 |
| O2-lab | Infuse 9500 | 126298.9 | 277661.3 | 632838.8 | 2.20 |

## 9. Thermal properties of fibers characterization via Differential Scanning Calorimetry

The thermal properties of the fibers are tested by using differential scanning calorimetry (DSC). DSC thermograms were obtained using a calorimeter model Discovery 2500 from TA Instruments. Heating-cooling cycles have been run from -40 °C to +180 °C in a nitrogen atmosphere. The nitrogen purging rate was 50 ml/min. The heating and cooling rates were 10 °C/min.

Before running DSC experiments, industrially-spun yarns were cleaned with Deionized (DI) water and isopropyl alcohol (IPA). As a typical cleaning procedure, around 20 cm of yarn was placed in 1000 ml Pyrex beaker with DI water - IPA mixture (ratio is 50/50 vol/vol) at room temperature for 24 h, then sonicated for 30 min in pure IPA and then for 30 min in DI water using a sonication bath (Branson model). The cleaned sample is dried in a VWR vacuum oven at room temperature.

The shape of the melting endothermic dips and the positions of their maxima on DSC thermograms depend on the fiber molecular structure, such as chain branching or volume ratio of crystalline to amorphous regions. For example, an H1-ind fiber with high content of crystallizable phase exhibits a sharp melting peak and high value of specific melting enthalpy (~190 J/g). In turn, a stretchable O1-lab fiber with a high amorphous phase content exhibits broad melting endothermic dips with lower value of specific melting enthalpy (~45 J/g).

The values of the peak melting temperatures and crystallinity values of PE-based fibers are summarized in **Supplementary Table S8**. The weight crystallinity of fibers was estimated as:

$$X_{DSC}(\%) = \frac{\Delta H_m - \Delta H_{cc}}{\Delta H_0} \times 100$$

where $\Delta H_0$ is the specific enthalpy of fusion ($\Delta H_0$ is 290 J/g for OBC [18,19] and 294 J/g for PE [7,20]), $\Delta H_m$ is the measured specific melting enthalpy of fiber from the DSC thermogram, and enthalpy of cold crystallization $\Delta H_{cc}$ is 0 in this work. Higher crystallinity fibers exhibit higher peak melting temperatures, which is in good agreement with prior literature data for semicrystalline polymers [18–28].

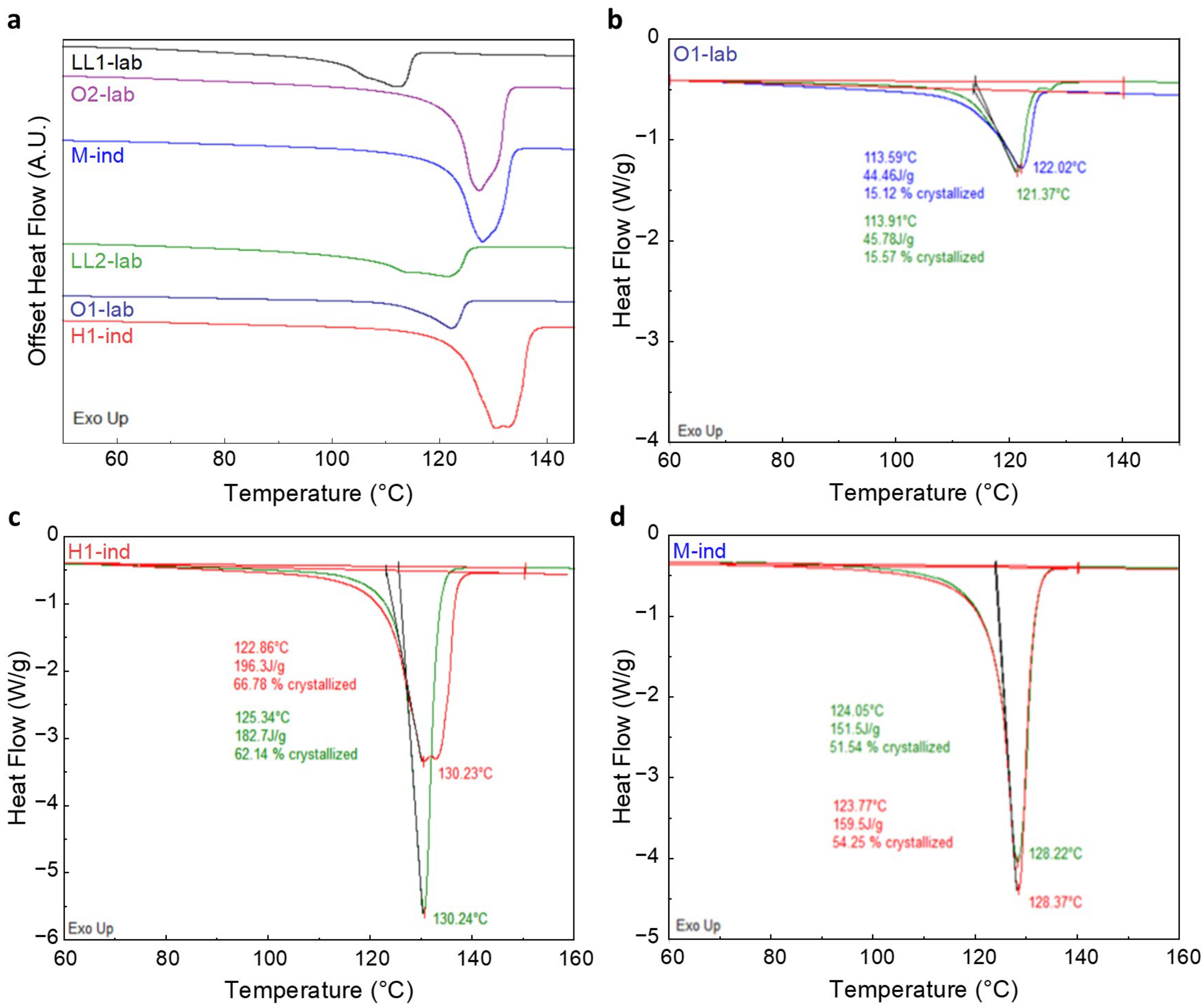

**Supplementary Figure S8.** (a) DSC thermograms reveal the melting temperatures and crystallinity levels of the PE fibers; the curves are vertically offset along the Heat Flow axis for visual clarity, and this offset is applied for display purposes only. (b-d) Examples of the analysis of DSC thermograms of O1-lab (*b*), H1-ind (*c*) and M-ind (*d*) fibers.

**Supplementary Table S8. Fiber properties extracted from the DSC thermograms**

| Label | Melting temperature, $T_m$ (°C) | Crystallinity (%) |
|---|---|---|
| O1-lab | 122 | 14 |
| O2-lab | 127 | 61 |
| LL1-lab | 112 | 27 |
| LL2-lab | 121 | 38 |
| LL3-ind[7] | 122 | 52 |
| M-lab | 128 | 56 |
| M-ind | 128 | 53 |
| H1-ind | 130 | 65 |

## 10. Fiber structure characterization by wide-angle X-ray scattering (WAXS)

The fiber crystallinity and orientation of crystalline domains along the fiber direction were evaluated using a wide-angle X-ray scattering (WAXS) system with Rigaku 002 microfocus Cu Kα X-ray source and a DECTRIS PILATUS 300K detector. X-ray experiments were performed in a vacuum chamber to minimize radiation damage and scattering from air. The fiber axis was perpendicular to the incident X-ray beam in each experiment.

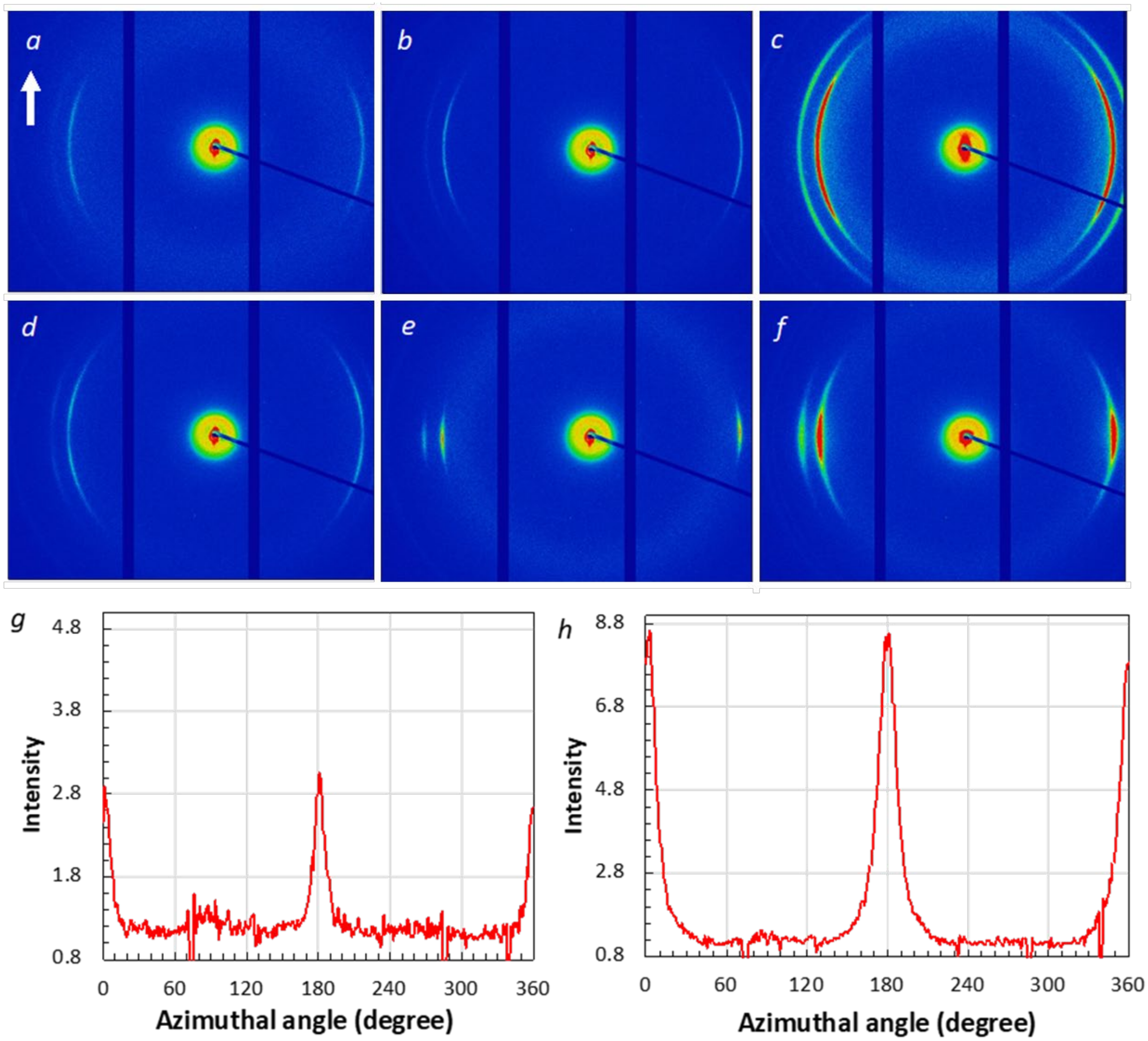

**Supplementary Figure S9.** (a-f) 2D WAXS patterns for LL1-lab (a), O2-lab (b), M-lab (c), LL2-lab (d), O1-lab (e), and H1-ind (f) fibers. (g-h) 1D azimuthal spectra of O1-lab (g) and H1-ind (h) fibers.

The X-ray diffraction profiles were extracted via a traditional narrow rectangular filter (size of this filter defined by width of diffraction arc [29]) and analyzed in an open-source software Fityk 1.3.1 for data processing and curve fitting [30,31]. After background subtraction, each diffraction curve was decomposed into an amorphous halo and the crystalline peaks. The diffraction peaks were fitted by Gaussian distributions.

The fiber crystallinity is calculated via deconvolution of multiple peaks in the X-ray diffraction spectra. The crystallinity has been calculated as the ratio of the sum of the areas under deconvoluted crystalline (110), (200), (210) peaks and the total area, which includes both the crystalline peaks and the amorphous halo [7].

The degree of orientation of crystalline domains along the fiber axis was estimated by analyzing the azimuthal distribution of the scattering intensity and calculation of Herman's orientation factor (HOF) [19,25]. The average domain orientation can be calculated as:

$$\langle \cos^2\phi \rangle = \frac{\int_0^{\pi/2} I(\phi)\cos^2\phi \sin\phi \, d\phi}{\int_0^{\pi/2} I(\phi)\sin\phi \, d\phi}$$

where $\phi$ is the azimuthal angle between the axis of the molecular segment and of the fiber and $I(\phi)$ is the scattering intensity at that angle. The Herman's orientation factor is defined as:

$$f = \frac{3\langle \cos^2\phi \rangle - 1}{2}$$

Function $f = -0.5$ if the orientation is perpendicular to the fiber direction, a value of $f = 1$ if it is parallel to the fiber direction, and $f = 0$ when the orientation is random.

**Supplementary Table S9. Fiber properties extracted from the WAXS spectra**

| Label | Crystallinity (%) | HOF |
|---|---|---|
| O1-lab | 29 | 0.000 |
| O2-lab | 65 | 0.020 |
| LL1-lab | 28 | 0.010 |
| LL2-lab | 40 | 0.051 |
| LL3-ind[7] | 52 | 0.267 |
| M-lab | 65 | 0.17 |
| M-ind | 64 | 0.176 |
| H1-ind | 74 | 0.109 |

## 11. Repeated thermo-mechanical recycling stability of HDPE/OBC blended fibers

To evaluate the thermo-mechanical recyclability of the polyethylene-family fiber system, the representative H-O1 blend (90 vol% HDPE / 10 vol% OBC, Exp-22-BT-1405 + Infuse 9100) was subjected to repeated melt-reprocessing and respinning cycles. In each cycle, the previously spun monofilament was pelletized, re-melt-compounded at 20 rpm, and re-extruded into a monofilament on the Xplore microcompounder at 175 °C. Fiber properties were characterized after 0, 1, 3, 5, and 10 cycles.

The blend remained continuously spinnable across all investigated cycles, with no extrusion interruption or melt-flow instability observed. Extrusion temperature, screw speed, and spool torque remained within the operating range used for the as-spun (non-recycled) fiber, indicating that the melt processability of the H-O1 system is preserved over repeated reprocessing. Fiber diameter variations across cycles reflect intentional adjustments of the take-up speed and drawing rather than degradation-induced flow instability.

**Supplementary Table S10.** Processing parameters of H-O1 monofilaments fabricated by repeated melt-reprocessing and respinning on the Xplore microcompounder. Note that the as-spun H-O1 used as the starting point of the recycling sequence (0 cycles) was spun separately from the H-O1 in **Tables S2b and S5**.

| Label | Additional Passes Through Extruder | Fiber Diameter (um) | Draw Ratio | Extrusion/Compound Temperature (deg C) | Spool Speed (m/min) | Screw Speed (rpm) | Spool Torque (Nm) |
|---|---|---|---|---|---|---|---|
| H-O1 | 0 | 199.3 | ~ 15 | 175 | 13.00 | 1 | 40 |
| H-O1-R1 | 1 | 316.7 | ~ 10 | 175 | 18.00 | 1 | 30 |
| H-O1-R3 | 3 | 132.0 | ~ 23 | 175 | 13.00 | 1 | 40 |
| H-O1-R5 | 5 | 116.7 | ~ 26 | 175 | 13.00 | 1 | 40 |
| H-O1-R10 | 10 | 143.0 | ~ 27 | 175 | 13.00 | 1 | 40 |

* All melt compounded at 20 rpm before extrusion.

The mechanical properties of the recycled fibers are summarized in **Supplementary Table S11**. Young's modulus remained stable across all recycling cycles, varying within ~515–660 MPa over ten reprocessing iterations, with the 10-cycle fiber (~660 MPa) comparable to or slightly higher than the as-spun fiber (~552 MPa). This indicates that the bulk stiffness of the polyethylene matrix is preserved through repeated melt-reprocessing. The tensile strength and elongation at break exhibited cycle-to-cycle variation consistent with the sensitivity of these quantities to the applied draw ratio, but no monotonic deterioration was observed over the investigated range. Together, these results demonstrate that the H-O1 system can tolerate multiple thermo-mechanical recycling steps while retaining both melt processability and bulk mechanical integrity, supporting its compatibility with existing polyethylene recycling streams.

**Supplementary Table S11.** Mechanical properties of H-O1 monofilaments after repeated melt-reprocessing and respinning, measured by tensile testing (gauge length 54 mm, extension rate 15 mm/min)

| Label | Maximum Force, N | Ultimate Tensile strength, MPa | Elongation at break, % | Young's modulus, MPa | Specific strength, MPa·m$^3$/kg | Specific stiffness, MPa·m$^3$/kg |
|---|---|---|---|---|---|---|
| H-O1 | 4.58 | 29.40 | 345 | 552.1 | 0.031 | 0.583 |
| H-O1-R1 | 8.872 | 22.52 | 414 | 548.0 | 0.024 | 0.579 |
| H-O1-R3 | 4.014 | 58.67 | 193 | 546.7 | 0.062 | 0.577 |
| H-O1-R5 | 3.853 | 72.06 | 121 | 527.4 | 0.076 | 0.557 |
| H-O1-R10 | 3.707 | 46.15 | 265 | 658.0 | 0.049 | 0.695 |

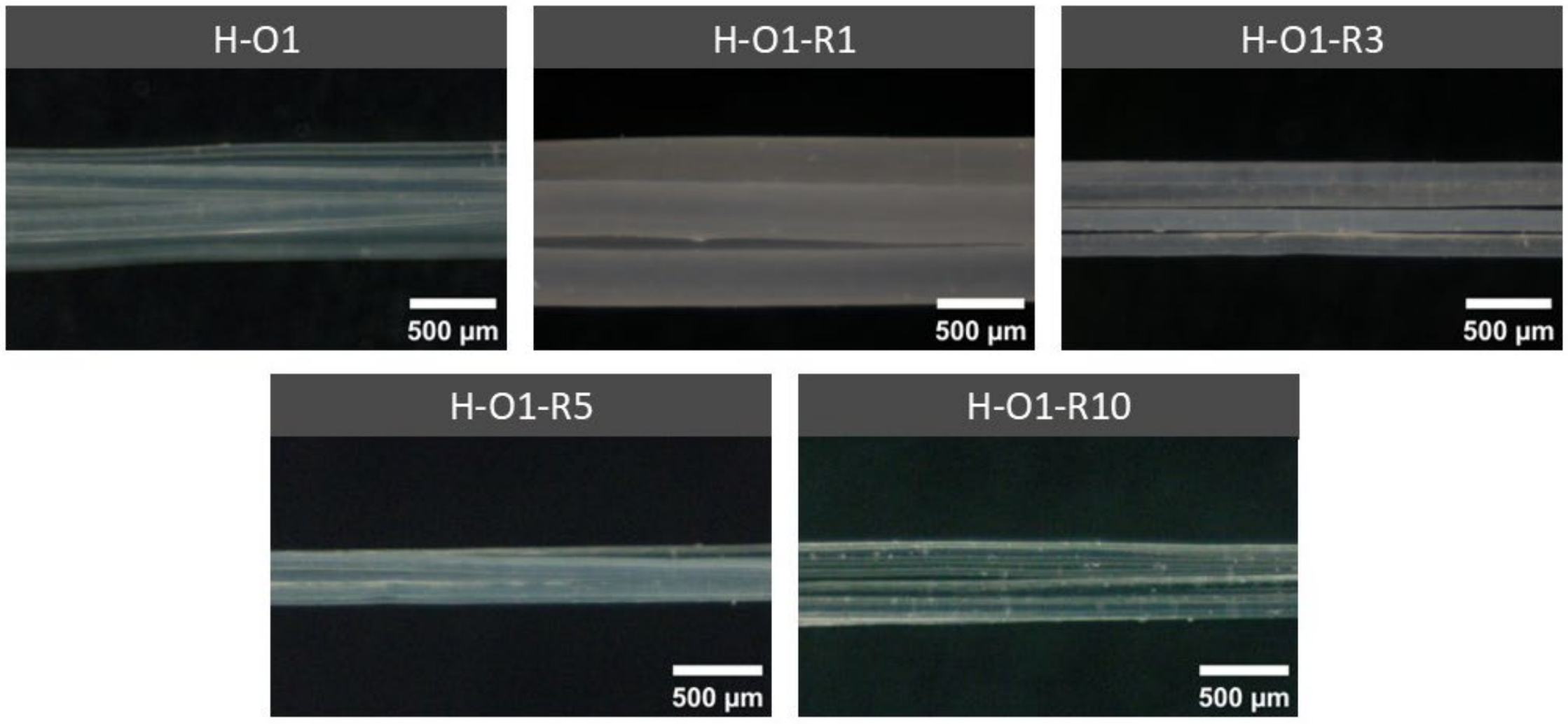

**Supplementary Figure S10.** Optical microscopy images of H-O1 fibers after 0, 1, 3, 5, and 10 additional thermo-mechanical recycling cycles. Scale bars: 500 µm.

## 12. Characterization of a commercial PET-spandex stretchable yarn

Commercial core-sheath hybrid yarn Sorbtek® used in this work was donated by FLEX LTD. The Sorbtek® core-sheath yarn is composed of a 150-denier, 34 filament polyester yarn wrapped around a 20-denier monofilament spandex core with a Z twist.

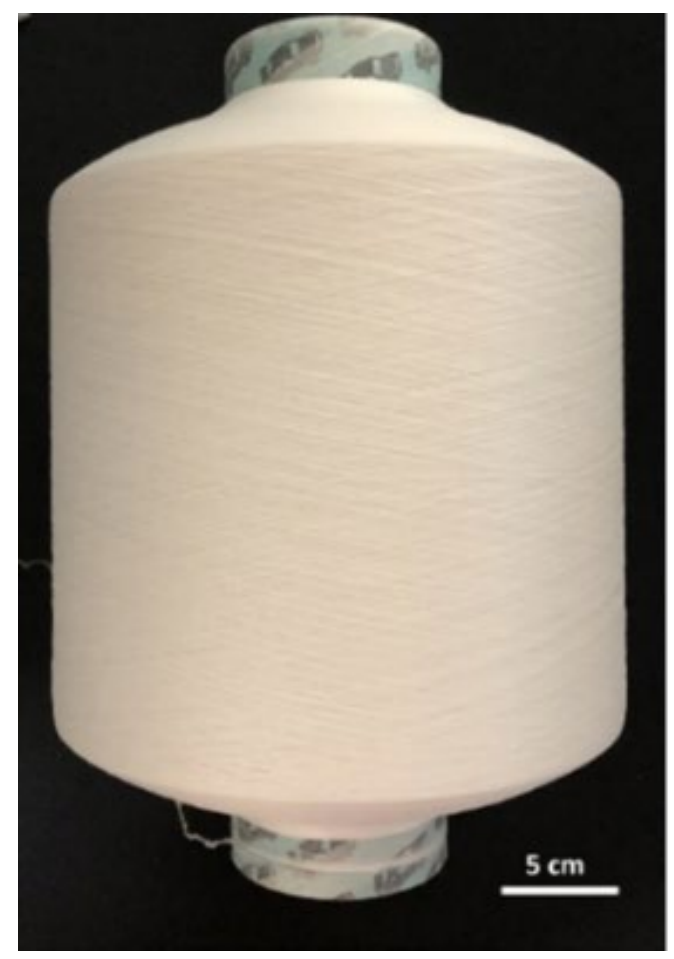

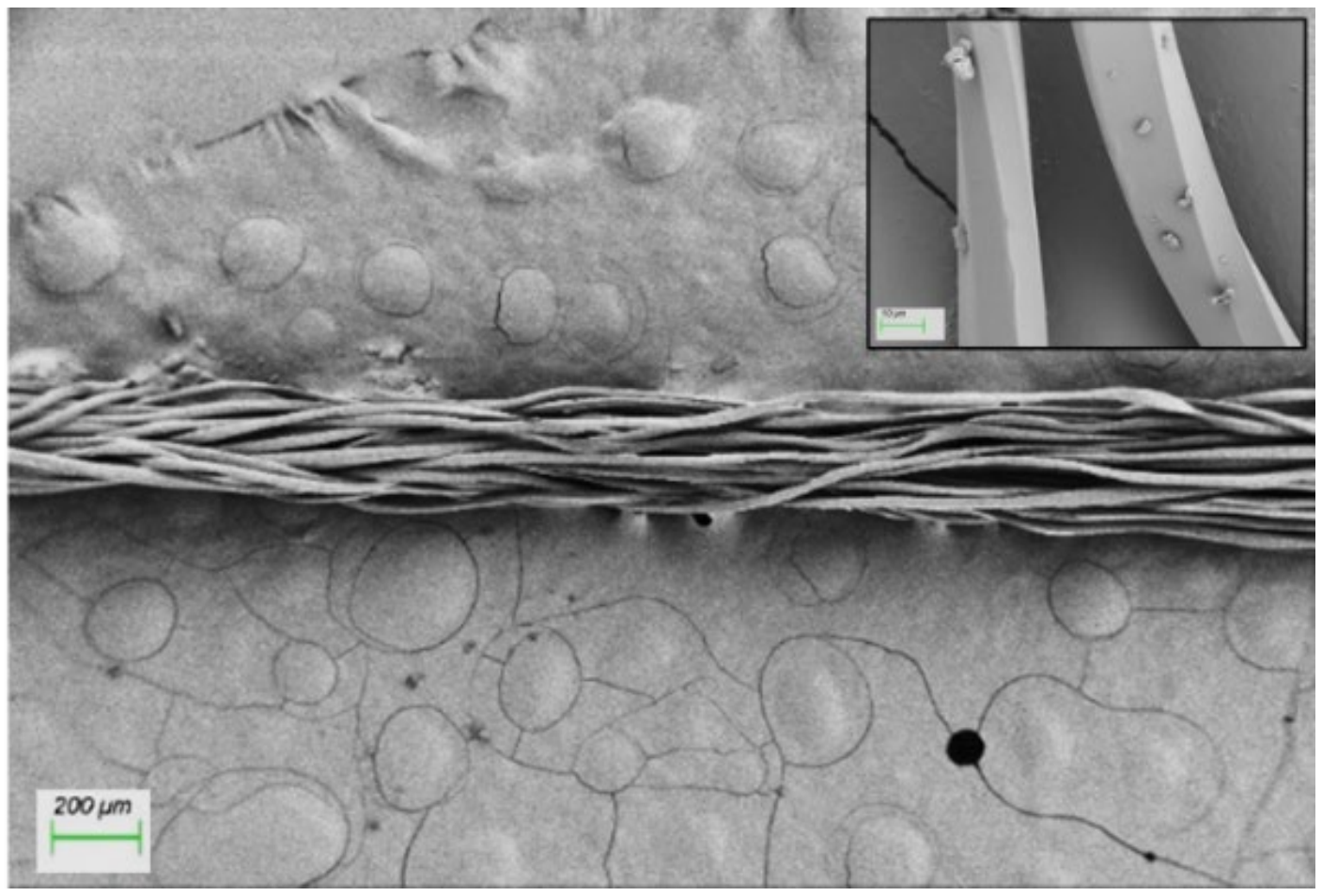

**Supplementary Figure S11.** Optical (left) and scanning electron microscopy (SEM, right) images of a commercial elastic Sorbtek® yarn. The inset shows an SEM image of polyester fibers comprising the sheath of the complex yarn. Scale bars are 200 micron for the SEM image of the yarn and 10 micron for the SEM image of fibers.

The morphology and structural characterization of the Sorbtek® yarn was performed by a high-resolution scanning electron microscope (Zeiss) operated at 3 kV with a secondary electron detector. For SEM imaging, the samples were coated with a layer of gold. The average fiber diameter was obtained from over 20 measurements performed on the SEM images with the help of the ImageJ software. We estimate that the PET filaments composing the sheath yarn have effective diameters of 22±5 micron, while the elastic spandex core has an effective diameter of 304±40 micron (see **Supplementary Figure S11**).

## 13. Fabrication of polyethylene-based stretchable yarns

Stretchable polyethylene-based yarns were fabricated by co-twisting an OBC core with multi-filament HDPE or MDPE yarns on the cone-to-cone multi-function DirecTwist 2C6 twisting machine (AGTEKS company). The OBC core with a diameter of 73 µm (34 denier) was stretched to about twice its original length, and then twisted with a multi-filament yarn, which was fed without tension, to form the core-sheath yarn shown in **Figs. 4a, d** of the manuscript. Upon releasing the loading tension, the core filament retracted to its original length, leading to the formation of a helical sheath around the core.

## 14. Validation of industrial-scale spinnability of OBC-homopolymer PE blended yarns

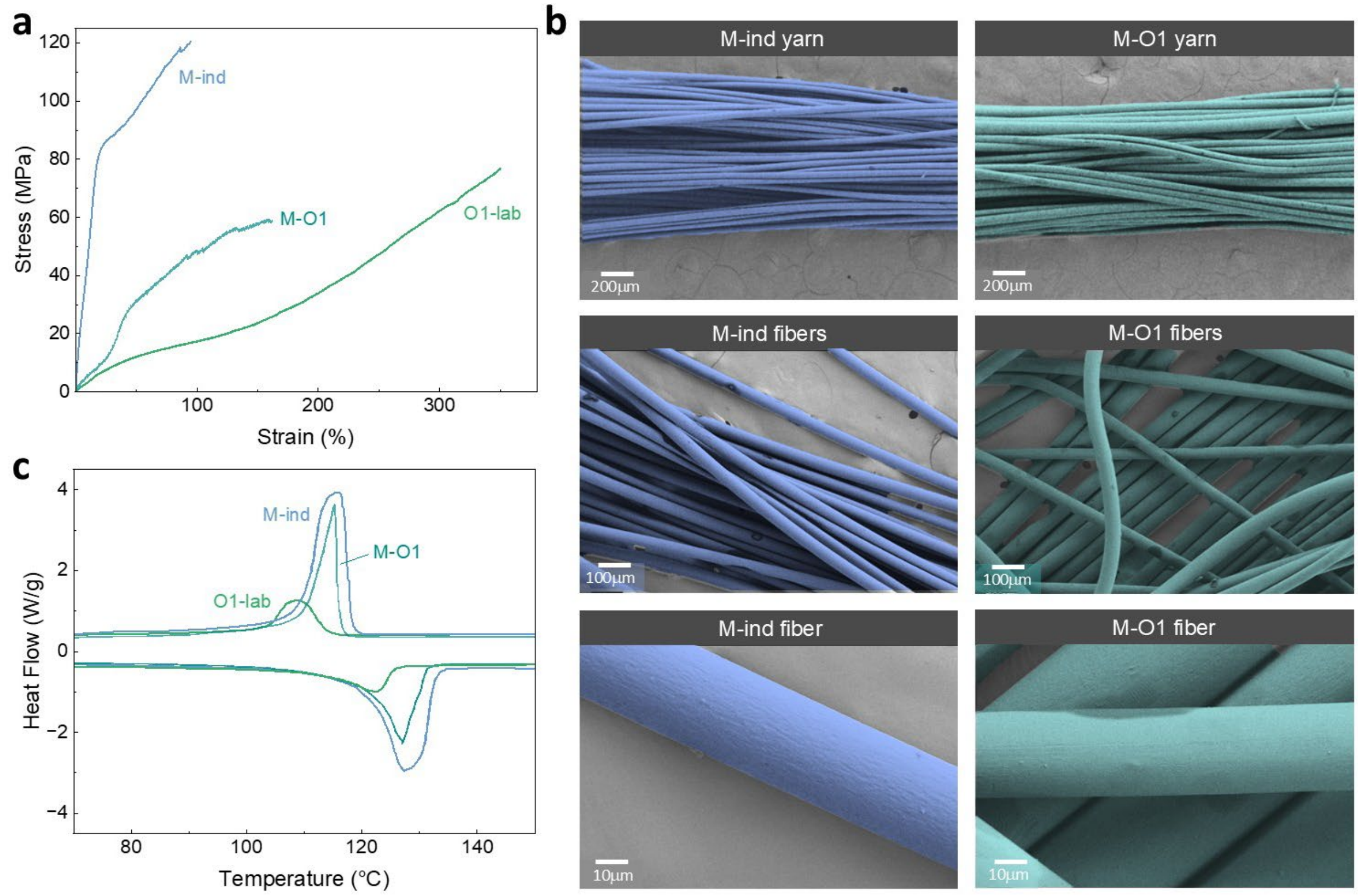

**Supplementary Figure S12.** (a) Tensile stress-strain curves for fibers spun on an extruder from M and O1 resins and their 1:1 blend, where the 1:1 composition represents an extreme case selected to evaluate the compatibility of PE-OBC blends and their spinnability on an industrial extruder. (b) SEM micrographs at multiple magnifications reveal the morphology of the yarns and individual fibers. (c) DSC thermograms of mono-material and blended fibers are consistent with co-crystallization of the two blended components.